\documentclass{elsarticle}
\usepackage{graphicx}
\usepackage{natbib}
\usepackage{float}
\usepackage{subfig}
\usepackage{setspace}
\usepackage{color}
\usepackage{multirow}
\usepackage{amsmath}
\DeclareMathOperator{\tr}{tr}

\graphicspath{{./Figures/}}

\title{Feed-Forward Control of a Backward-Facing Step Flow}
\author{N. Gautier and J.-L. Aider}
\address{Laboratoire de Physique et M\'ecanique des Milieux H\'et\'erog\`enes (PMMH), UMR7636 CNRS, \\ \'Ecole Sup\'erieure de Physique et Chimie Industrielles de la ville de Paris\\ 10 rue Vauquelin,  75005 Paris, France}

\begin{document}
\maketitle
\begin{abstract}
Closed-loop control of an amplifier flow is experimentally investigated. A feed-forward algorithm is implemented to control the flow downstream a backward-facing step. Upstream and downstream data are extracted from real-time velocity fields to compute an ARMAX model used to effect actuation. This work, done at Reynolds number 430 investigates the practical feasibility of this approach which has shown great promise in a recent numerical study by \cite{Sipp2012}. The linear nature of the regime is checked, 2D upstream perturbations are introduced, and the degree to which the flow can be controlled is quantified. The resulting actuation is able to effectively reduce downstream energy levels and fluctuations. The limitations and difficulties of applying such an approach to an experiment are also emphasized.
\end{abstract}

\section{Introduction}
Closed-loop flow control is of major academic and industrial interest. At the interface of control theory and fluid mechanics it is pertinent to many engineering domains, such as aeronautics and combustion. It can be used to reduce aerodynamic drag of an automobile or an airplane, increase combustion efficiency, or enhance mixing. Control of amplifier flows like boundary layers, mixing layers, jets or separated flows is particularly relevant and challenging. Indeed, amplifier flows are globally stable, however convective instabilities will amplify disturbances while being advected downstream (\cite{Jacquin2008,Marquet2011}). Incoming perturbations are likely to be amplified to the point where they disrupt the entire flow. Nullifying these disturbances before they can be amplified by the flow is a great challenge for the flow control community (\cite{Henningson2001}). Typically when considering a laminar amplifier flow,  the control objective can be to inhibit the transition to turbulence. Examples of such flows abound, a much-studied amplifier flow being that over a backward-facing step (BFS) (\cite{Barkley2002,Blackburn2008}) which presents an unsteady region of convective instability. Another example is that of the flow over a cavity, used for studying the control of global instabilities (\cite{Rowley2006}).
\\
Control of amplifier flows has been the subject of much research (\cite{Sipp2012,Belson2013}). A control law can be computed using one of two ways. One possibility is to compute the model using beforehand knowledge of the physics of the flow (\cite{Barbagallo2010}). When derived directly from the Navier-Stokes equations these models are of very high order and require reduction before they can be used in a realistic setting.  Model reduction is still a rich and very active research field, see \cite{Efe2003,Rowley2004,Akervik2007}.  In some cases a physical analysis of the flow can yield simple models leading to efficient control laws as shown in \cite{Pastoor2008,Gautier2013frequency}. The second option is system identification as suggested by \cite{Bagheri2009}. In this case, the flow is probed until a model can be derived from its responses. This approach is data based: it seeks to build an input-output model for the flow from empirical observations. Such an approach has been applied with success to the control of the recirculation bubble behind a BFS, see \cite{King2005,King2007}.
\\\\
The BFS is considered as a benchmark geometry for the study of amplifier flows: separation is imposed by a sharp edge creating a strong shear layer susceptible to Kelvin-Helmholtz instability. Upstream perturbations are amplified in the shear layer leading to significant downstream disturbances. This flow has been extensively studied  both numerically and experimentally (\cite{Armaly1983,Le1997,JLA2004,JLA2007}).
\\\\
The principle of feed-forward control is to act on the flow upon detection of an event as opposed to the more common feed-back control where one reacts to an event. Feed-forward algorithms have been successfully used in flow control in numerical simulations (\cite{Belson2013}). Recently \cite{Sipp2012} have shown the effectiveness of a feed-forward algorithm computed using an Auto-Regressive Moving-Average Exogenous model (ARMAX) to capture the relevant dynamics of the flow. The resulting control law leads to reduced energy levels and fluctuations. The aim of this work is to determine the feasibility and robustness of this approach in an experimental setting.

\section{Experimental Setup}
\subsection{Water tunnel}
Experiments were carried out in a hydrodynamic channel in which the flow is driven by gravity.
The flow is stabilized by divergent and convergent sections separated by honeycombs.  The quality of the main stream can be quantified in terms of flow uniformity and turbulence intensity.  The standard deviation $\sigma$ is computed for the highest free stream velocity featured in our experimental set-up. We obtain $\sigma = 0.059$~cm.s$^{-1}$ which corresponds to turbulence levels of $\frac{\sigma}{U_{\infty}}=0.0023$. For the present experiment the flow velocity is $U_{\infty} =$~2.1~cm.s$^{-1}$  giving a Reynolds number based on step height $Re_h=\frac{U_{\infty}h}{\nu} = 430$. Following the assumptions of \cite{Sipp2012} Reynolds number was chosen to ensure a sub-critical linear 2D flow.

\begin{figure}
\centering
\includegraphics[width=0.75\textwidth]{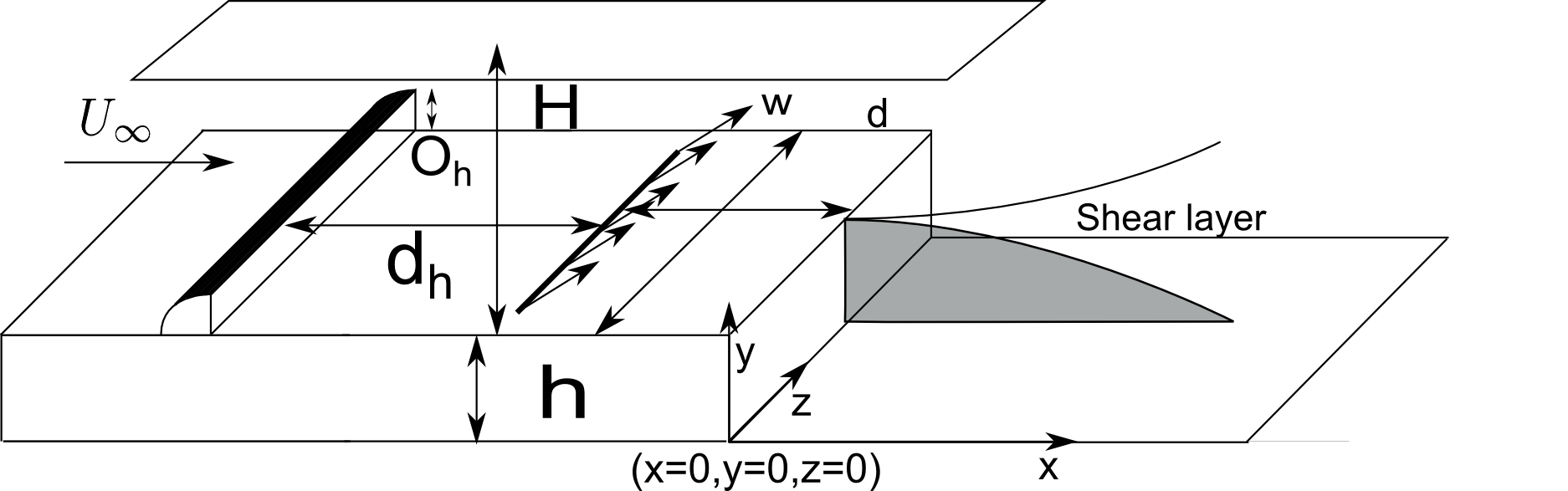}
\caption{Sketch of the BFS geometry and definition of the main parameters.}
\label{fig:dimensions}
\end{figure}

\subsection{Backward-Facing Step geometry and upstream perturbation}
The BFS geometry and the main geometric parameters are shown in figure~\ref{fig:dimensions}. BFS height is $h=1.5$~cm. Channel height is $H=7$~cm for a channel width $w=15$~cm. The vertical expansion ratio  is $A_y = \frac{H}{h+H} = 0.82$ and the span-wise aspect ratio is $A_z=\frac{w}{h+H}=1.76$. The injection slot is located $d / h =2$ upstream of the step edge. 
\\
The principle of the method described in \cite{Sipp2012} is to devise an input-output model for the flow based on experimental data. This model is used to compute actuation aimed at negating incoming upstream noise, thereby preventing its amplification. Because our sensor is 2D in the symmetry plane and our actuator can only deliver span-wise homogeneous actuation, a 2D upstream perturbation is required for effective control. As shown in figure~\ref{fig:dimensions} a 2D obstacle with a rounded leading edge of height $O_h=0.8$~cm has been placed at $d_h=15$~cm  upstream from jet injection (12~h from the step edge). Because of the low Reynolds number the flow stays 2D.

\begin{figure}
\centering
a) \includegraphics[width=0.95\textwidth]{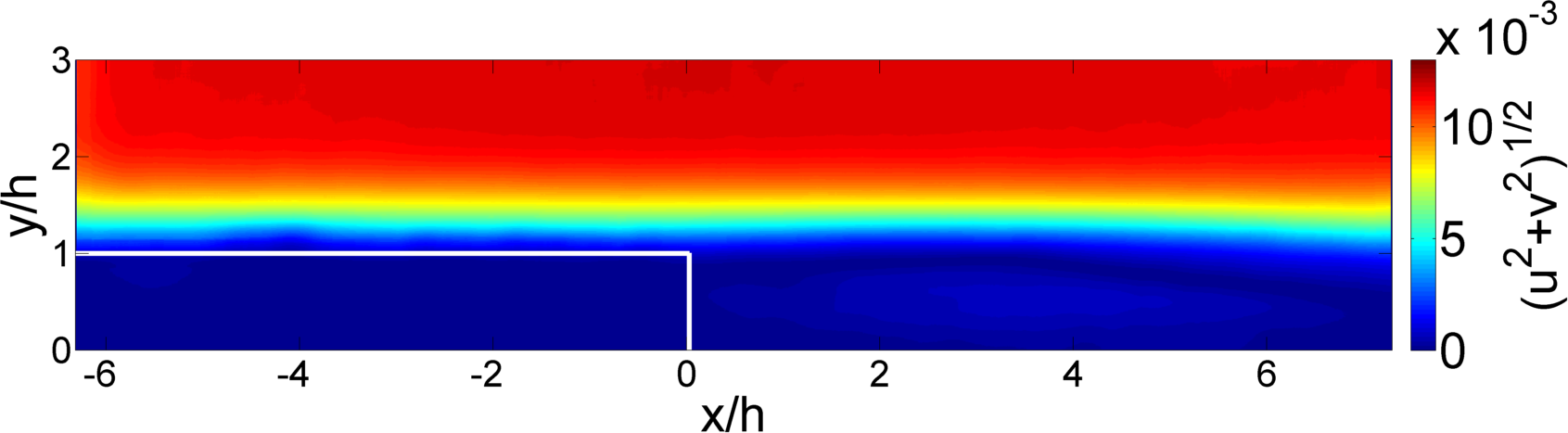}
\\
b) \includegraphics[width=0.95\textwidth]{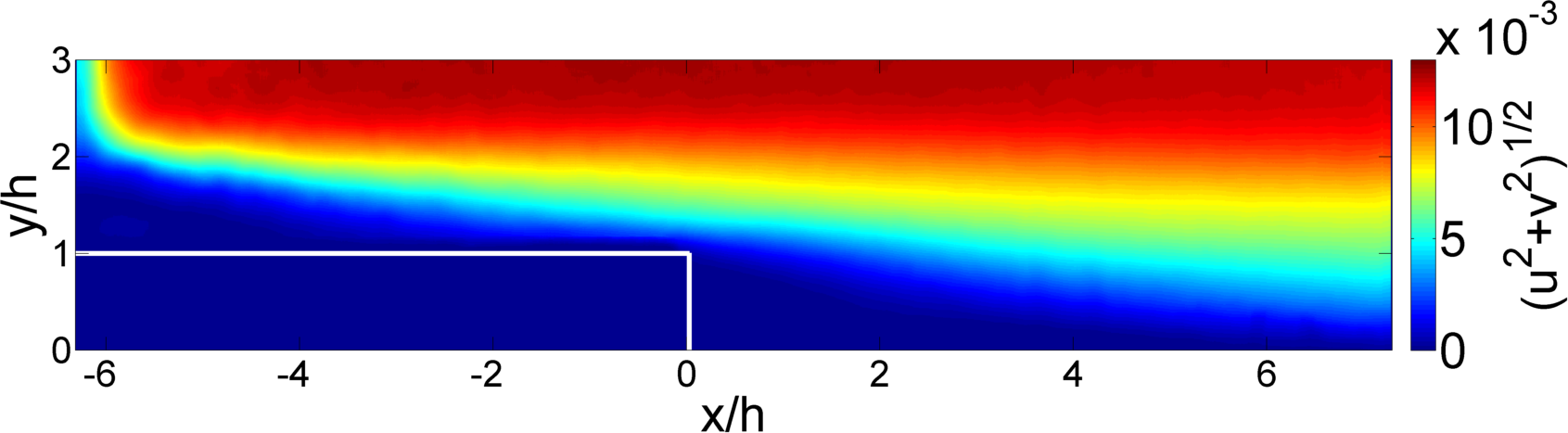}
\\
c) \includegraphics[width=0.95\textwidth]{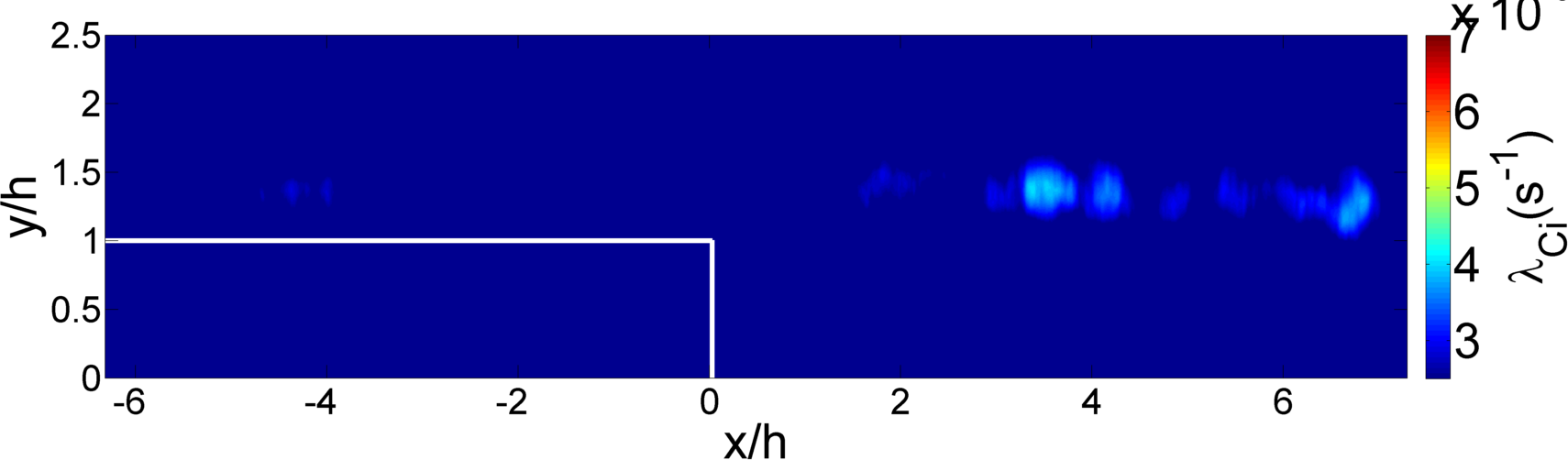}
\\
d) \includegraphics[width=0.95\textwidth]{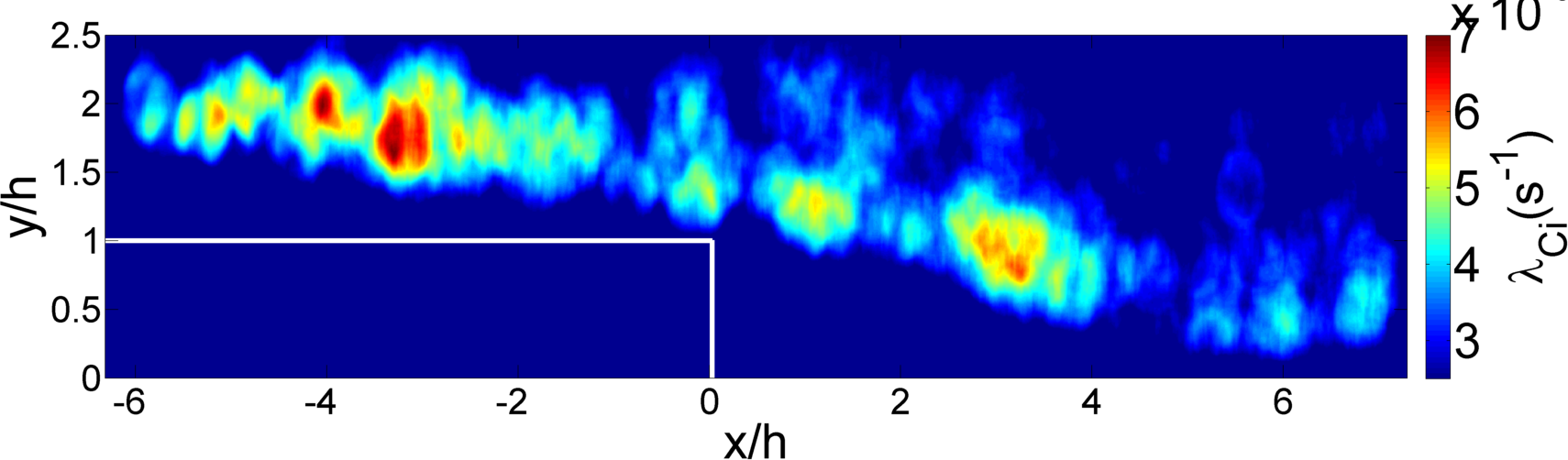}
\caption{Mean velocity magnitude contour fields for the flow with (b) and without obstacles (a). Instantaneous snapshots showing contours of $\lambda_{Ci}$ for the flow with (d) and without obstacle (c).}
\label{fig:rolls_ups}
\end{figure}

\subsection{Sensor: 2D real-time velocity fields computations}
The sensors used as inputs for the closed-loop experiments are visual sensors, i.e. regions of the 2D PIV (Particle Image Velocimetry) velocity fields measured jn the symmetry plane as shown in figure \ref{fig:window}. The flow is seeded with 20~$\mu$m  neutrally buoyant polyamid seeding particles.  They are illuminated by a laser sheet created by a 2W continuous laser beam operating at $\lambda$~=~532~nm.  Images of the vertical symmetry plane are recorded using a Basler acA 2000-340km 8bit CMOS camera. Velocity field computations are run on a Gforce GTX 580 graphics card.
The algorithm used to compute the velocity fields is based on a Lukas-Kanade optical flow algorithm called FOLKI developed by \cite{Champagnat2005}. Its offline and online accuracy has been demonstrated and detailed by \cite{Plyer2011,Gautier2013OF}.  Furthermore this acquisition method was successfully used in \cite{Leclaire2012,Gautier2013control}.
The size of the velocity fields is $17.2\times4.6$~cm$^2$. They are computed every $\delta_t=20ms$, for a sampling frequency $F_s=25Hz$.

\subsection{Uncontrolled flow}
The swirling strength criterium $\lambda_{Ci}$ is an effective way of detecting vortices in 2D velocity fields introduced and improved by \cite{Chong1990,Zhou1999}. For 2D data the swirling criterium is defined as  $\lambda_{Ci}=\frac{1}{2}\sqrt{4 \det(\nabla \bf{u})- \tr(\nabla \bf{u})^2}$ (when this quantity is real).\\
 Figures~\ref{fig:rolls_ups} a) and b) show the mean velocity amplitude fields for the uncontrolled flow with and without obstacle. Figures~\ref{fig:rolls_ups} c) and d) show $\lambda_{Ci}$ snapshots of the uncontrolled flow with and without the obstacle, highlighting the perturbations caused by the upstream obstacle. Figure~\ref{fig:rolls_ups} d) shows the steady stream of vortices created by the obstacle interacting with the recirculation. Quantitatively, $\lambda_{Ci}$ is an order of magnitude higher than for the flow without obstacle.
\\The boundary layer thickness at the step edge for the flow with and without obstacles are $\delta =$~1.34h and $\delta =$~1.73h respectively.

\begin{figure}
\centering
\includegraphics[width=0.65\textwidth]{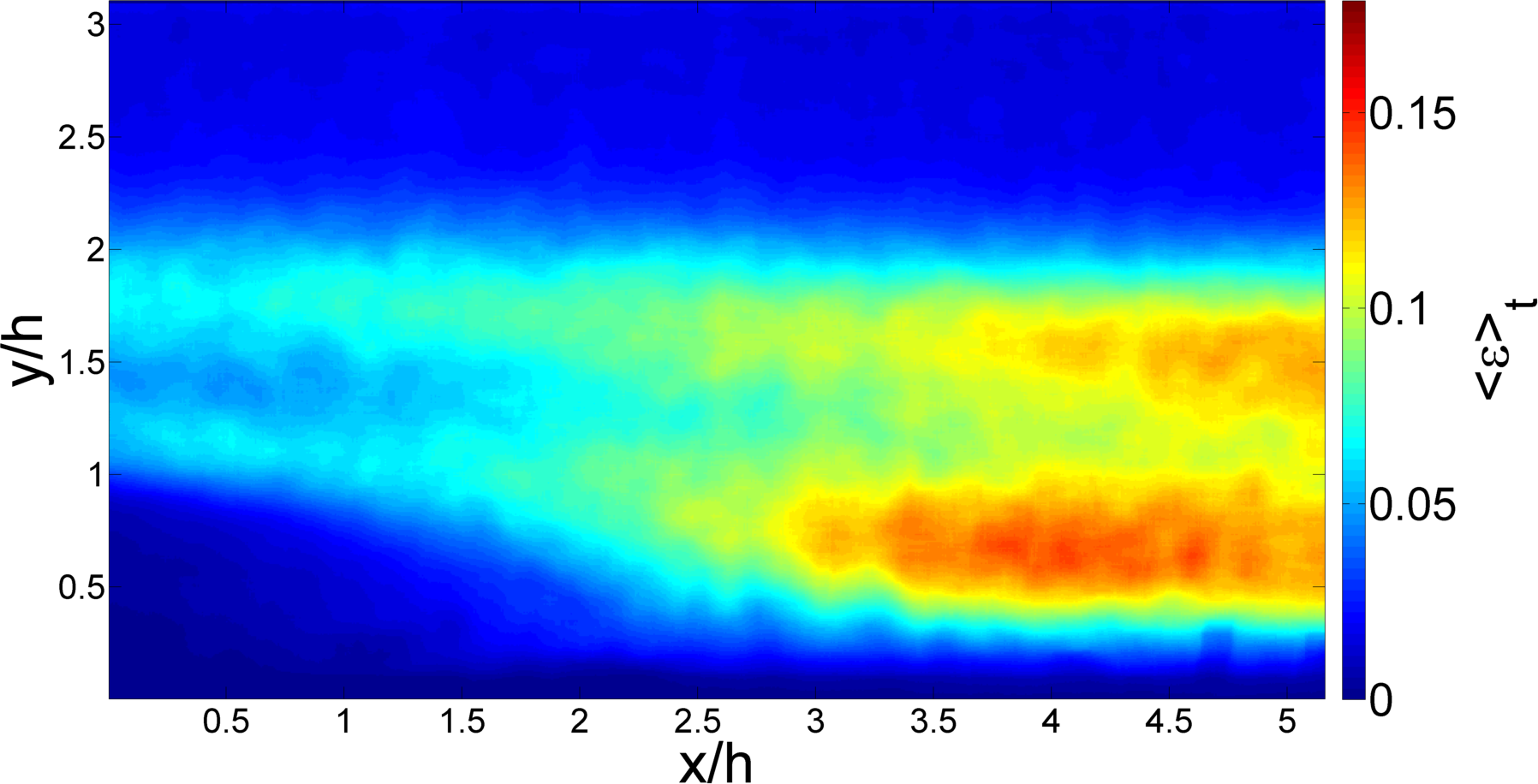}
\caption{TKE field $<\epsilon (x, y) >_t$ downstream the BFS with the upstream obstacle.}
\label{fig:mean_TKE_un}
\end{figure}

The turbulent kinetic energy (TKE) is defined as $\epsilon(x,y,t)=\frac{1}{2}(u'(x,y,t)^2+v'(x,y,t)^2)$, where u', v' are longitudinal and vertical velocity fluctuations. The figure~\ref{fig:mean_TKE_un} shows the time-averaged TKE field $<\epsilon(x,y)>_t$ downstream of the step for the case with the upstream obstacle. The field exhibits two regions of high TKE. The lower region corresponds to the recirculation bubble. The upper region corresponds to residual  perturbations induced by vortices shed by the upstream obstacle. 

\subsection{Actuation}
\begin{figure}
\centering
\includegraphics[width=0.55\textwidth]{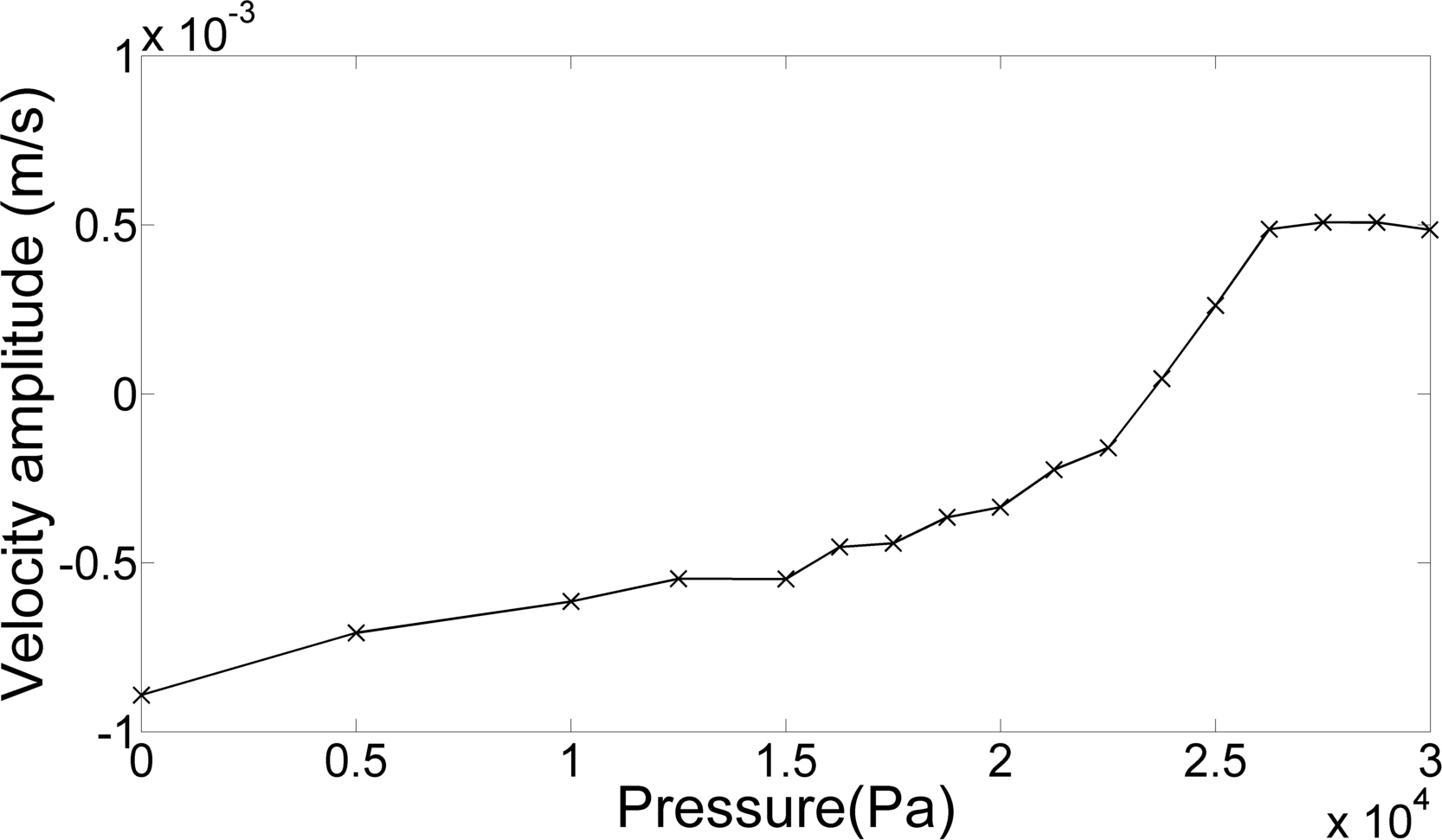}
\caption{Pressure to velocity transfer function}
\label{fig:pressure_to_amplitude}
\end{figure}

\cite{Sipp2012} actuation is a gaussian flow sink/source placed above the step, which is not experimentally feasible. In our case, actuation is provided by a flush slot jet,  0.1~cm long and 9~cm wide. This actuation has been chose to obtain a perturbation as homogeneous along the span-wise direction as possible. The jet angle to the wall is 45$^o$. The slot is located 3~cm (2h) upstream the step edge (figure~\ref{fig:dimensions}). Jet flow is induced using water from a pressurized tank. It enters a plenum and goes through a volume of glass beads designed to homogenize the incoming flow. Jet amplitude is controlled by changing tank pressure. Because channel pressure is higher than atmospheric pressure this allows us to provide both blowing and suction. The convection time from jet injection to measurement area is 2 s ($<0.5Hz$). The maximum actuation frequency $f_a$ is about 1Hz which is sufficient for these experiments.  \\
The control law output velocities, these are converted into pressure commands using the transfer function described in figure \ref{fig:pressure_to_amplitude}.

\section{ARMAX model}
%\subsection{ARMAX model}
\subsection{Introduction}
An ARMAX model is used because it can be derived from experimental data, \cite{Garnier2010}. Furthermore it has been shown by \cite{Sipp2012} that it is particularly well adapted at modeling the BFS flow when in the linear regime. 

\begin{figure}
\centering
\includegraphics[width=1.0\textwidth]{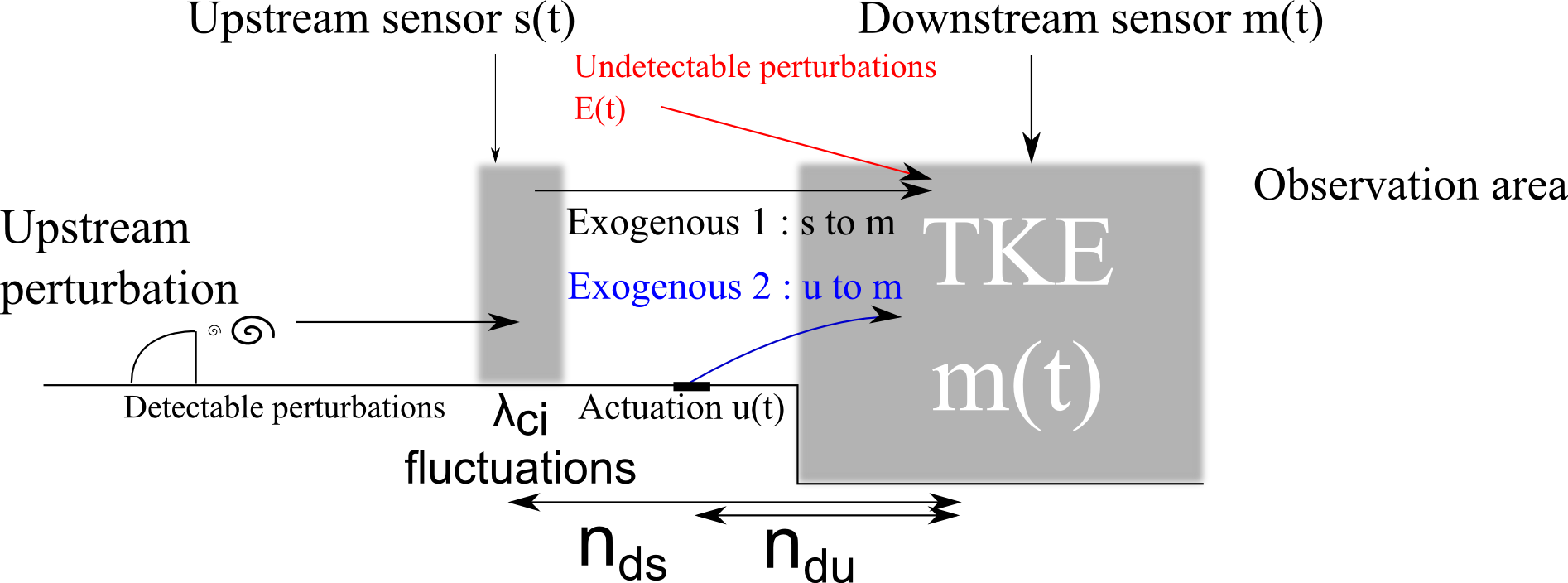}
\caption{Schematic description of the main terms used in the ARMAX model.}
\label{fig:window}
\end{figure}

Two exogenous inputs $s(t),u(t)$ and one output $m(t)$ are used. The first exogenous input $s(t)$ measures fluctuations of spatially averaged $\lambda_{Ci}$ (small grey area on figure \ref{fig:window}). Such a sensor is well suited to the detection of upstream vortices created by the obstacle. The second exogenous input is jet flow velocity $u(t)$. 
\\
Output $m(t)$ is a measure of TKE fluctuations in the recirculation region. The control objective is to negate the incoming perturbations created by the obstacle in order to reduce overall downstream TKE fluctuations.  TKE is averaged over the whole downstream velocity field (large grey area on figure \ref{fig:window}):

\begin{gather}\label{eq:m}
m(t) = \frac{\int \epsilon(x,y,t) dx dy}{\int dx dy}
\end{gather} 

Following \cite{Sipp2012} the equation for the model is defined in eq \ref{eq:ARMAX}:

\begin{gather}\label{eq:ARMAX}
\underbrace{m(t)+\sum_{k=1}^{n_a} a_k m(t-k)}_{auto-regressive}=\underbrace{\sum_{k=n_{du}}^{n_{du}+n_{bu}} b_k^u u(t-k)}_{exogenous\:1}+\underbrace{\sum_{k=n_{ds}}^{n_{ds}+n_{bs}} b_k^s s(t-k)}_{exogenous\:2}+E(t)\\
E(t)=\underbrace{\sum_{k=n_{1}}^{n_{c}} c_k e(t-k)}_{moving\:average}+e(t)\nonumber
\end{gather}

\begin{figure}
\centering
\includegraphics[width=0.5\textwidth]{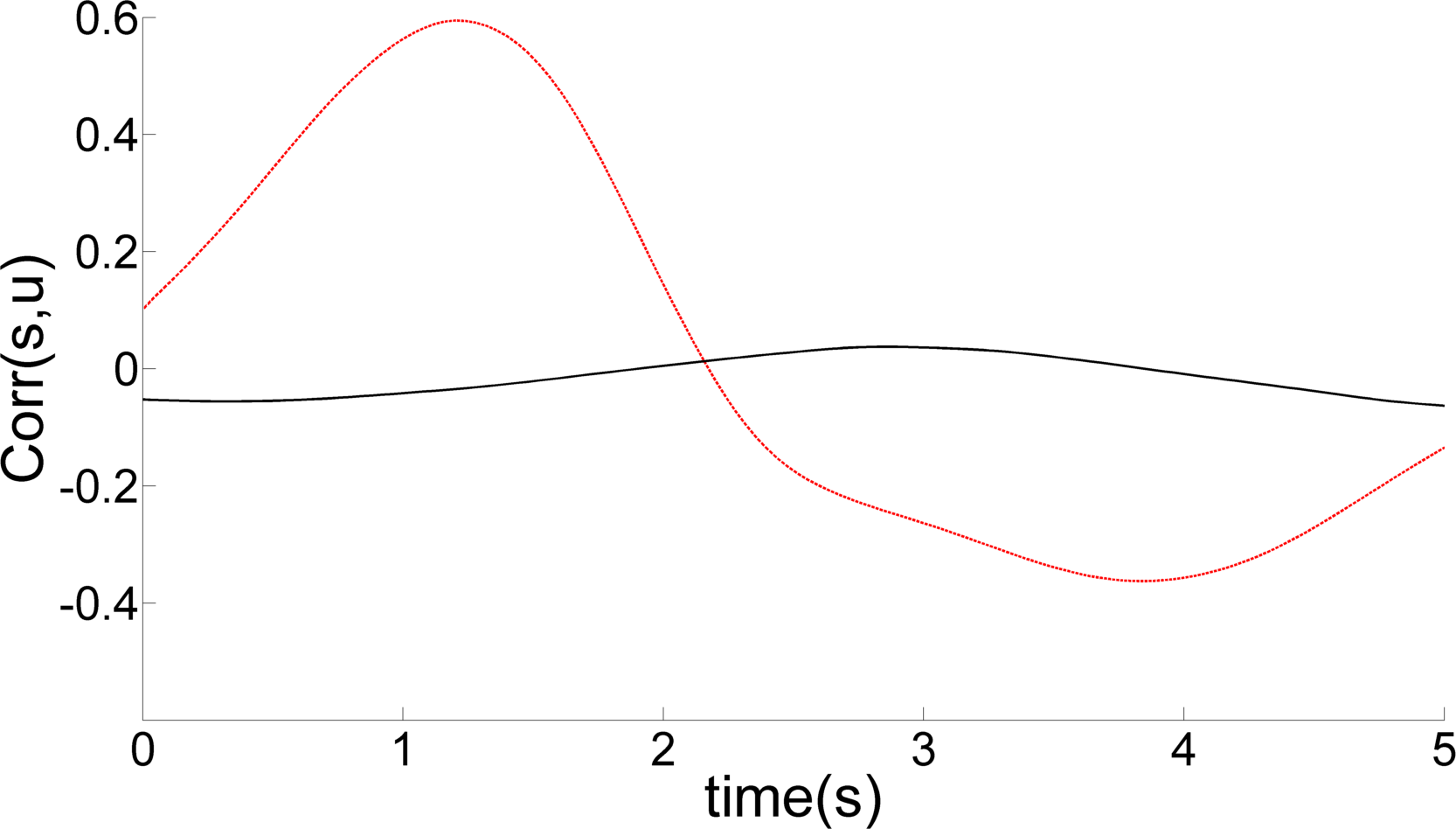}
\caption{Cross correlation functions between $s(t)$ and $u(t)$ for the calibration case (black) and a case with interference between the upstream sensor and the actuator (dotted red). }
\label{fig:corr_u_s}
\end{figure}

To achieve feed-forward control, the effects of upstream sensing $s(t)$ and actuation $u(t)$ on the output $m(t)$ must be quantified. For a pure feed-forward control, upstream estimation should be independent of actuation, see \cite{Semeraro2013}. During control, $u(t)$ is a function of $s(t)$. For our experimental setup we found that interference between actuation and the upstream sensor causes the control algorithm to saturate actuation. To avoid this effect, an inclined jet has been used instead of a wall normal jet. Moreover since $s(t)$ only measures the presence of vortices it is weakly affected by downstream actuation compared to vertical velocity for example. Special care must be given to lower actuation amplitude as much as possible so that it does not affect the upstream sensor. Figure \ref{fig:corr_u_s} shows the cross correlation function between $s(t)$ and $u(t)$ for two cases: the calibration case and a case where there is interference (jet amplitude is too high) between the upstream sensor and the actuator. Interference results in high correlation between $s$ and $u$ whereas in our calibration case correlation is negligible. 

Coefficients $(a_k, b_k^u, b_k^s)$ are computed to minimize error $e(t)$ at all times. To calibrate the model the user must provide time series for both inputs and outputs, the longer the better. Values for $n_a,n_{du}, n_{bu}, n_{ds}, n_{bs}$ are tied to the physics of the flow and are determined by the user. These coefficients are linked to time delays in the flow system. The flow time history required for the model to work properly is given by $n_a.\delta_t$ (auto-regressive part). $n_{du}.\delta_t$ and $n_{ds}.\delta_t$ are the times required for the respective inputs to affect the output; they are linked to flow convective velocity. $n_{bu}.\delta_t$ and $n_{bs}.\delta_t$ represent input time scales. They correspond to the time during which upstream effects  impact the output signal. Finally $n_c$ is used to model noise and ensures robustness (\cite{Sipp2012}). This value is chosen iteratively, once all other coefficients have been fixed, to get the best possible fit between experimental data and model output.

\subsection{Model Computation}
\begin{figure}
\centering
a)\includegraphics[width=0.75\textwidth]{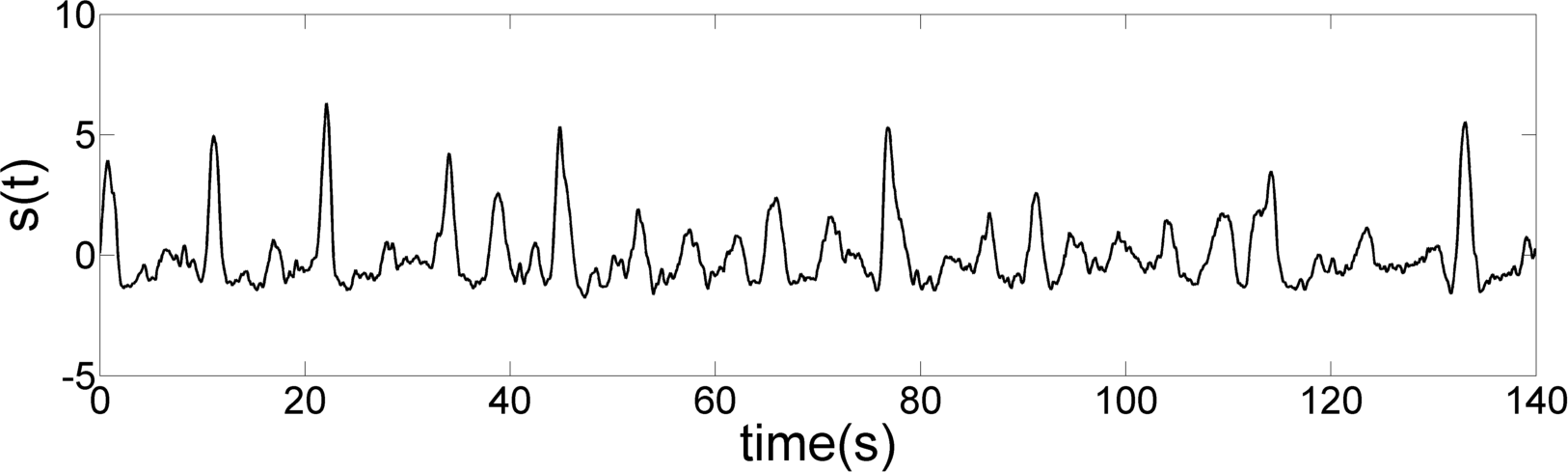}\\
b)\includegraphics[width=0.75\textwidth]{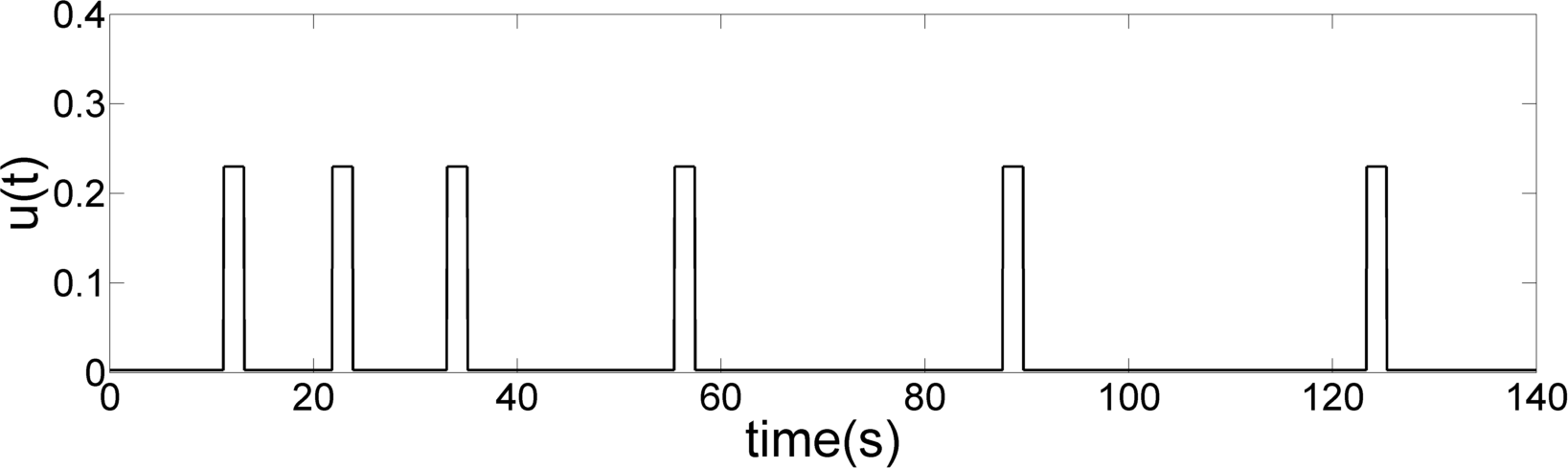}\\
c)\includegraphics[width=0.75\textwidth]{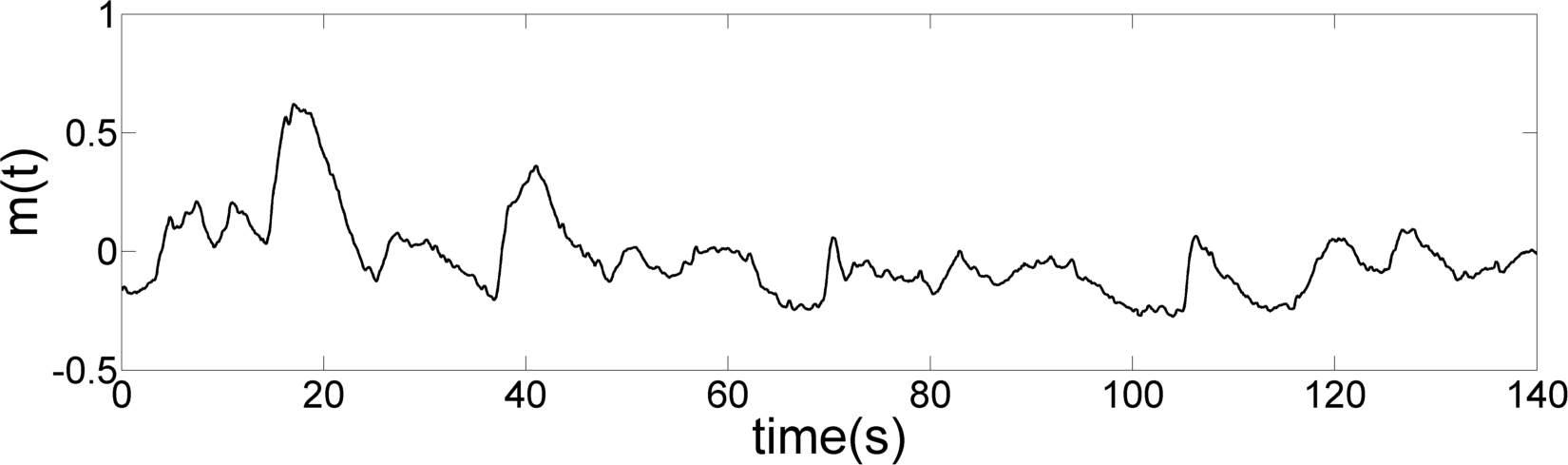}\\
\caption{Calibration time series. a) $s(t)$ captures the influence of upstream disturbances; b) $u(t)$ pseudo-random control law; c)  $m(t)$ spatially averaged downstream TKE.}
\label{fig:calibration_series}
\end{figure}

\begin{figure}
\centering
\includegraphics[width=0.75\textwidth]{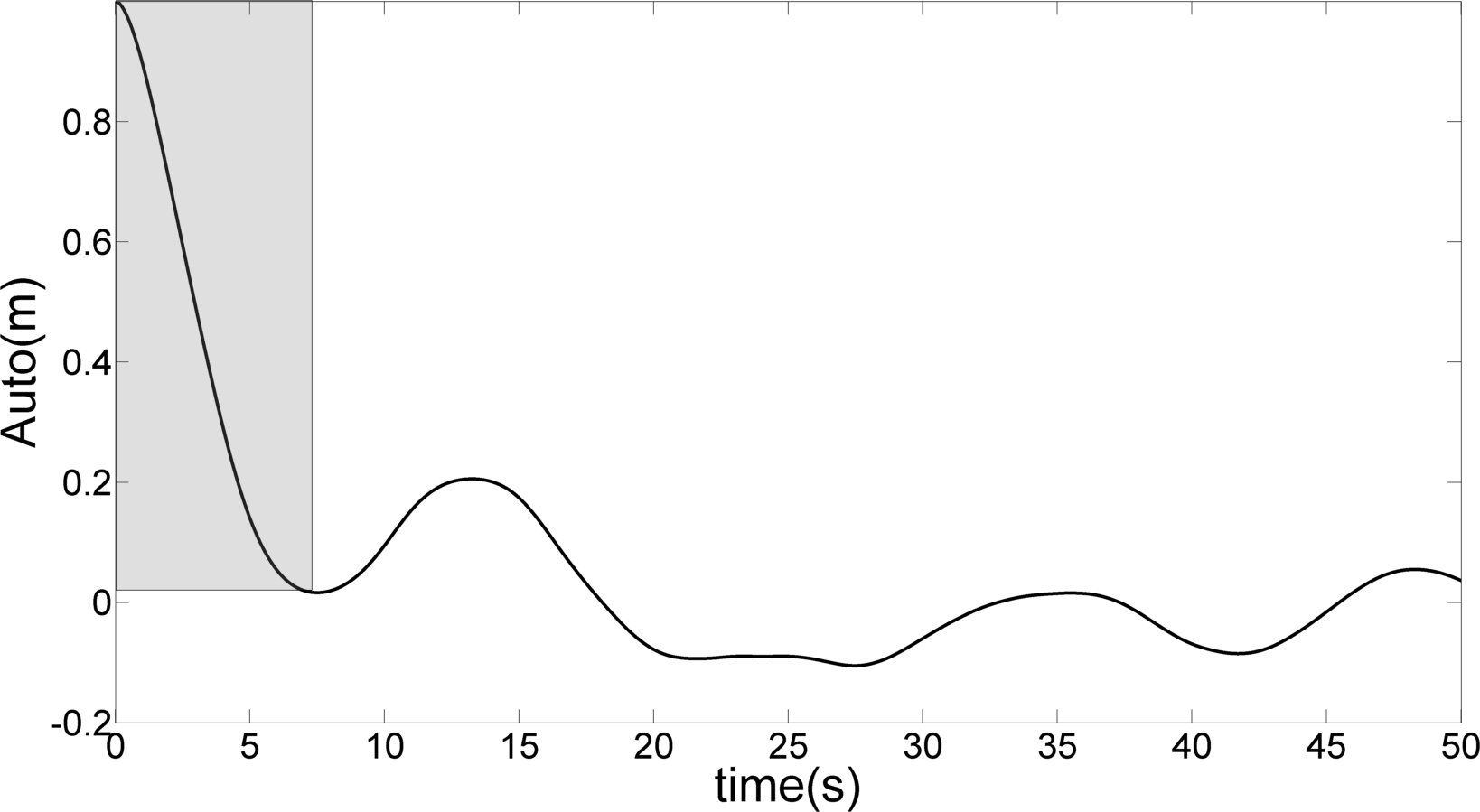}
\caption{Auto correlation function for $m(t)$.}
\label{fig:auto_m}
\end{figure}

Figure \ref{fig:calibration_series} shows a small segment of the calibration time series. The forcing law $u(t)$ used in these series is one of pseudo random pulses. Pulses are made to occur at random intervals, long enough for the effects of the previous pulse to have subsided before the next pulse. During these intervals the only input to the system is $s(t)$. This allows the effects of actuation and upstream perturbations to be computed using a single time series. Impulse amplitude for actuation $u(t)$ should be chosen such that it is high enough to affect the output $m(t)$ but low enough to avoid perturbations of the upstream output $s(t)$.
Calibration data were acquired over 25 minutes. Figure \ref{fig:auto_m} shows the auto correlation function for $m(t)$. A quasi-oscillatory behavior can be observed. It can be used to choose $n_a$ which is such that $n_a.\delta_t$ equals half the oscillatory period, as recommended by \cite{Sipp2012}.

\begin{figure}
\centering
\includegraphics[width=0.62\textwidth]{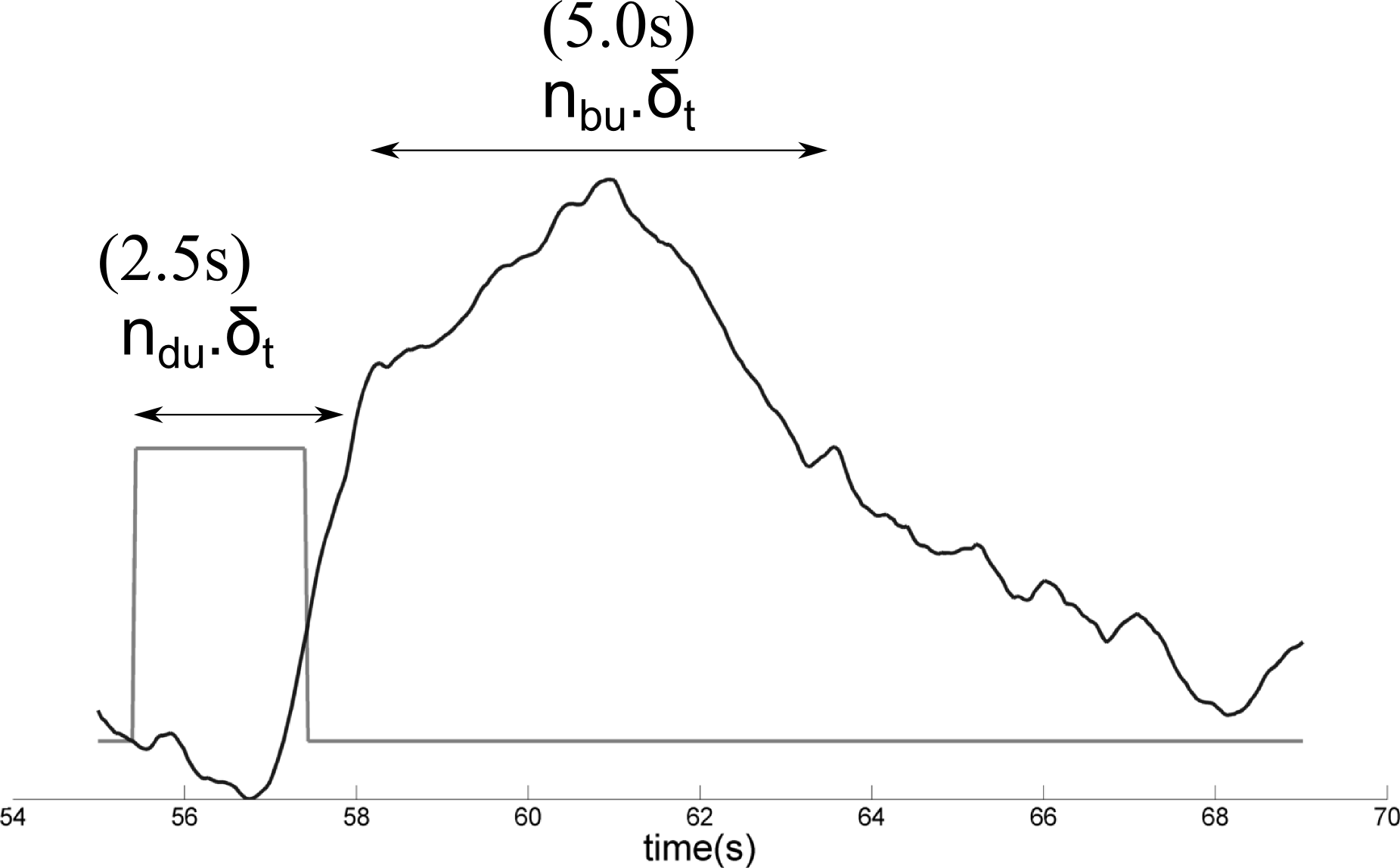}
\caption{Output impulse response}
\label{fig:exp_impulse_response}
\end{figure}

Figure~\ref{fig:exp_impulse_response} shows the response to an impulse, which can be used to evaluate the coefficients $n_{du},n_{bu},n_{ds},n_{bs}$. The time delay $t_{du}$=2.5s between the beginning of the actuation and the response gives $t_{du}=n_{du}.\delta_t$. The upstream sensor is located 3.5~cm upstream the jet injection.  Assuming perturbations travel at channel velocity, this implies a time delay of $t_{ds} \approx $~1.7~s for an upstream disturbance to affect the output, thus $n_{ds}.\delta_t= t_{du} + t_{ds}$. 
\\
Let $t_{bu}$ be the time during which an impulse in $u$ affects the output, as shown in figure \ref{fig:exp_impulse_response}, then $n_{bu}.\delta_t=t_{bu}$. Because the response to an impulse in $s$ is more difficult to distinguish we assume $n_{bs}=n_{bu}$. Finally $n_c$ is chosen after the other coefficients have been fixed in order to get the best  possible agreement between model and real outputs. Table~\ref{tab:coeff} summarizes the final coefficients used in the computation of the ARMAX model using the Matlab \textit{armax} function (\cite{Ljung1999}), it also shows the corresponding time delays and averages in seconds.

\begin{table}
\centering 
\begin{tabular}{c c c c c c }
$n_a$ \: & $n_{ds}$  \:& $n_{bs}$  \:& $n_{du}$  \:& $n_{bu}$ \: & $n_c$\\ \\
175  \:& 63  \:& 125  \:& 105 \: & 125  \:& 5 \\\\
7.0 s  \:& 2.5 s  \:& 5 s  \:& 4.2 s \: & 5 s  \:& 0.2 s
\end{tabular}
\caption{ARMAX coefficients}
\label{tab:coeff}
\end{table}

\begin{figure}
\centering
a) \includegraphics[width=0.65\textwidth]{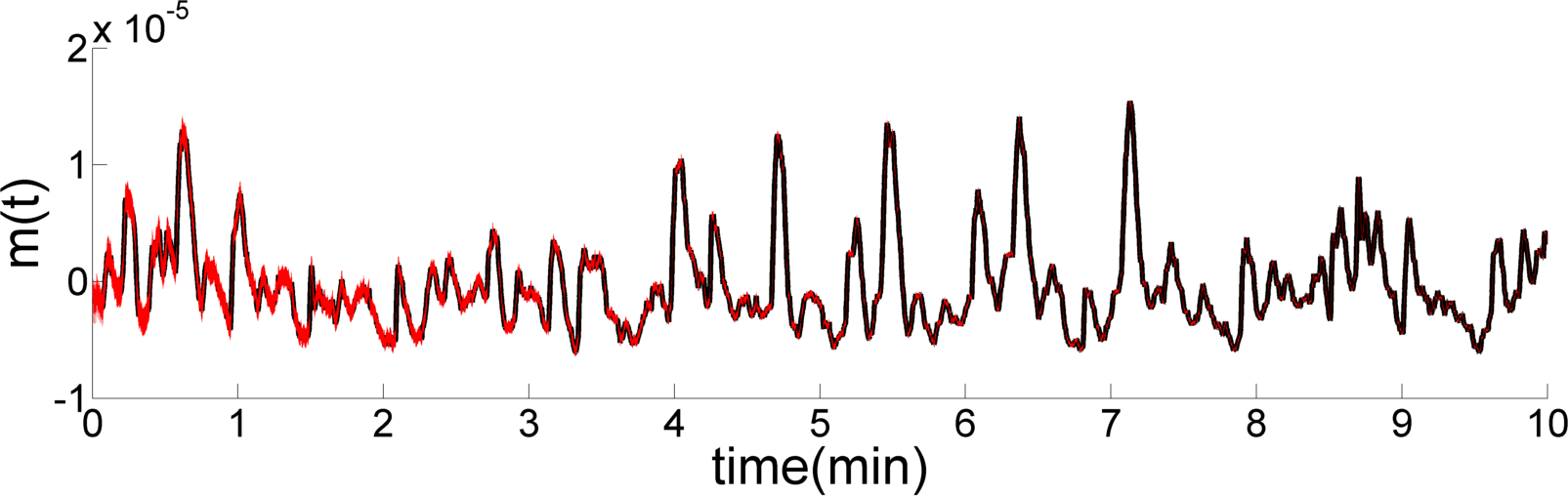}\\

b) \includegraphics[width=0.65\textwidth]{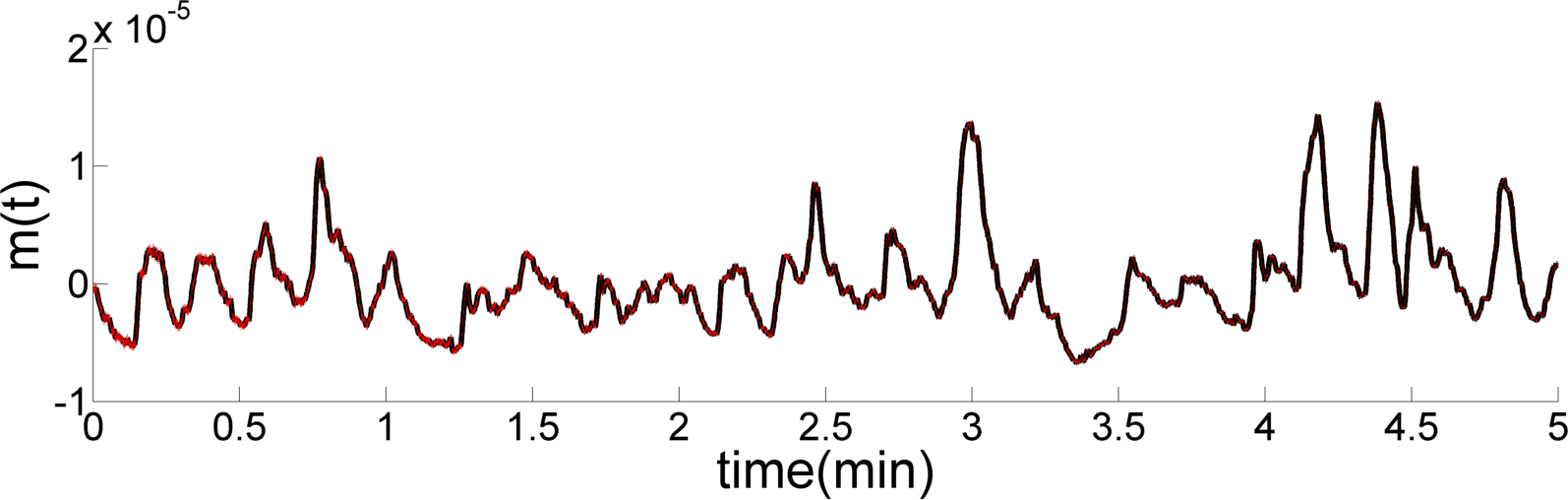}\\

\caption{a) Calibration data set, model performance (dotted red) compared to experimental results (in black). b) Validation data set, model performance (dotted red) compared to experimental results (in black).}
\label{fig:armax_output_validation}
\end{figure}

Figure \ref{fig:armax_output_validation}a compares ARMAX output to the source signal for the calibration series. Agreement is good at 96 \%. Figure~\ref{fig:armax_output_validation}b compares ARMAX output to the source signal for the validation series; agreement is slightly lower at 94 \%.

%\begin{figure}
%\centering
%\subfloat[Calibration data set, model performance (dotted black) compared to experimental results (in grey).]{\includegraphics[width=0.5\textwidth,height=0.2\textwidth]{armax_output_calibration}\label{fig:armax_output_calibration}}
%\subfloat[Validation data set, model performance (dotted black) compared to experimental results (in grey).]{\includegraphics[width=0.5\textwidth,height=0.2\textwidth]{armax_output_validation}\label{fig:armax_output_validation}}
%\end{figure}

\subsection{Linearity}
\begin{figure}
\centering
\includegraphics[width=0.5\textwidth]{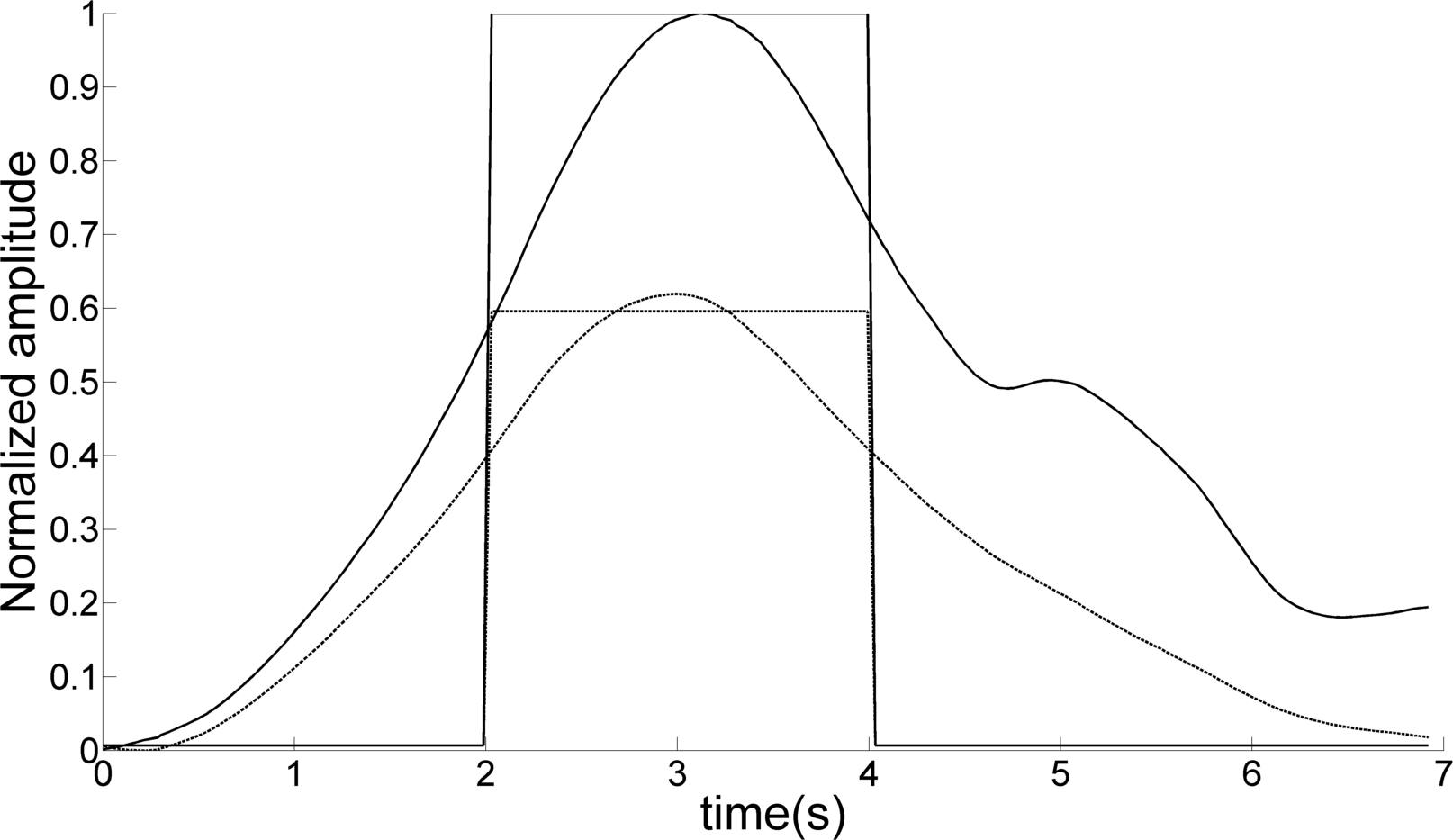}
\caption{Time evolution of the m(t) in response to a short impulse of different amplitudes (solid and dotted lines). The signals have been shifted in time to better highlight the linear nature of the response.}
\label{fig:linearity}
\end{figure}

A major underlying assumption of this approach is the linearity of the system. In our setup this was checked by imposing periodic pulsed forcing, with varying amplitudes. Figure \ref{fig:linearity} shows the phase averaged, spatially averaged TKE evolution in response to an actuation impulse. Impulse amplitude ratio is also given for comparison. A change in impulse amplitude leads to a proportional change in response amplitudes, confirming the linear behavior of the flow. Linearity was also checked when varying the size of the window where TKE is computed. Averaging over smaller windows, closer to the step, where non-linearities are weaker, did not improve the system linearity.

\section{Results}

\subsection{Control law}

\begin{figure}
\centering
a) \includegraphics[width=0.46\textwidth]{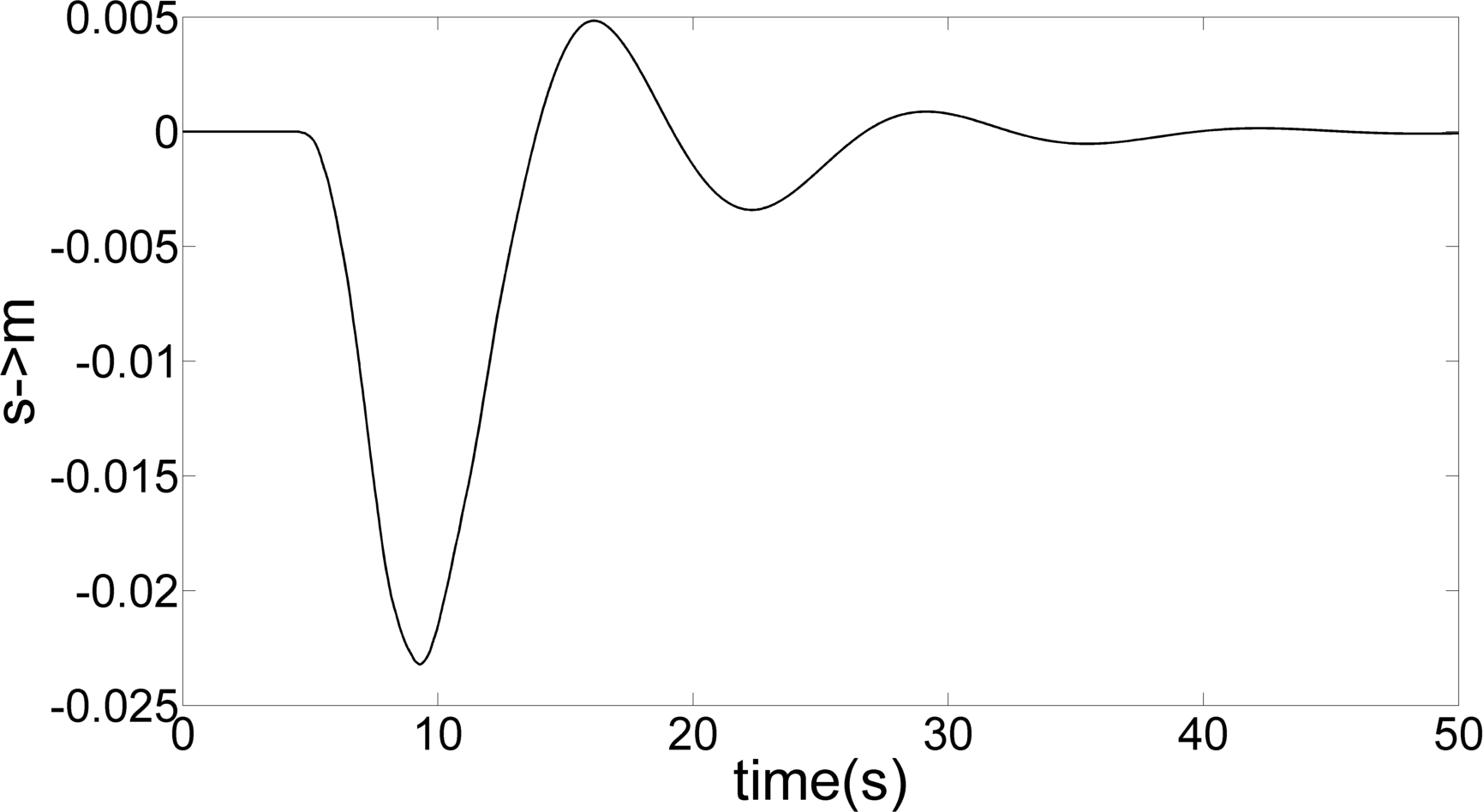}
b) \includegraphics[width=0.46\textwidth]{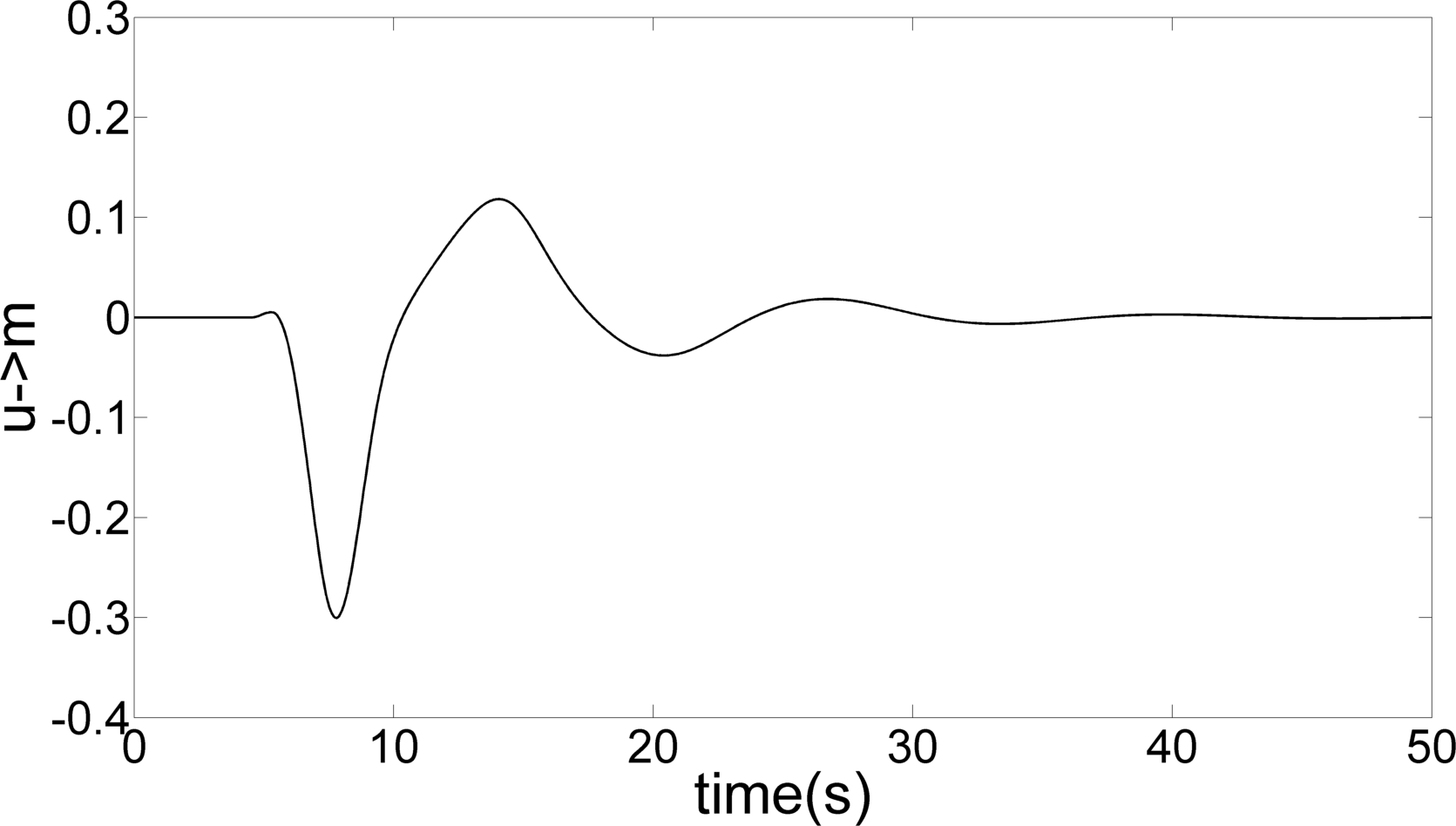}\\
\caption{a) ARMAX impulse response for exogenous input s(t). b) ARMAX impulse response for exogenous input u(t).}
\label{fig:impulse}
\end{figure}

Figures \ref{fig:impulse}a and \ref{fig:impulse}b show the impulse response for both exogenous inputs. These figures show impulse responses are qualitatively similar, however they differ in amplitude.
\\
Impulses responses can help determine if a model is "controllable" and whether or not the objective (negating TKE fluctuations) is \textit{a priori} attainable, which makes them an invaluable diagnostics tool. To achieve fluctuation suppression the control law suggested by \cite{Sipp2012} was computed.\\
Only perturbations detected in $s(t)$ can potentially be canceled out. Other sources of disturbance are not modeled and are ignored by the control law. Equation~\ref{eq:transfer}  illustrates how the output signal can be written as a combination of the input signals.
\begin{equation}
m(t)=\sum_{k=0}^{\infty} h_k^s s(t-k)+ h_k^u u(t-k)
\label{eq:transfer}
\end{equation}

The coefficients $h_k^s,h_k^u$ are obtained by computing the impulse response of the ARMAX model as described in equation~4.2 and \ref{eq:h} for an impulse response $s(t=0)=1,u(t=0)=1$.

\begin{gather}
\forall k \: m_{impulse \: s}(t=k)=h_k^s\: s(0)\\
\forall k \: m_{impulse \: u}(t=k)=h_k^u\:u(0)
\label{eq:h}
\end{gather}

These coefficients can be used to express $m(t)$ as a function of $s(t), u(t)$ as shown in equation \ref{eq:discrete}. Previously $s(t)$ was used to compute the model, here it is used as an input which allows us to compute $u(t)$.  This is done over 2000 time steps (T~=~80~s). 

\begin{equation}
M_T=H_u U^f+G_u U^P+G_s S^P
\label{eq:discrete}
\end{equation}
with 
\begin{multline*}
M_T=\begin{pmatrix}
  m_{t}  \\
 m_{t+1}  \\
  \vdots \\
m_{t+T}
 \end{pmatrix},
U^f=\begin{pmatrix}
  u_{t}  \\
 u_{t+1}  \\
  \vdots \\
u_{t+T}
 \end{pmatrix},
U^P=\begin{pmatrix}
  u_{t-1}  \\
 u_{t-2}  \\
  \vdots \\
u_{t-T}
 \end{pmatrix},
S^P=\begin{pmatrix}
  s_{t}  \\
 s_{t-1}  \\
  \vdots \\
s_{t-T}
 \end{pmatrix}\\
H_u=\begin{pmatrix}
  h_0^u &   &  &   \\
 h_1^u& h_0^u &   &   \\
  \cdots  & \cdots  & \ddots &   \\
  h_T^u & \cdots & \cdots & h_0^u
 \end{pmatrix},
G_u=
\begin{pmatrix}
  h_1^u & \cdots  &\cdots  & h_T^u   \\
 h_2^u& \cdots & h_T^2  & 0  \\
  \cdots  & \reflectbox{$\ddots$}  &  & 0  \\
  h_u^T & \cdots & \cdots & 0 \\
0 & 0 & 0 & 0
 \end{pmatrix},
G_s=
\begin{pmatrix}
  h_0^s & \cdots  &\cdots  & h_T^s   \\
 h_1^u& \cdots & h_T^s  & 0  \\
  \cdots  & \reflectbox{$\ddots$}  &  & 0  \\
  h_T^s & \cdots & \cdots & 0 \\
 \end{pmatrix}
\end{multline*}

Our goal is to find $U^f$ such that $M^T=0$ thus $U^f=(-H_u^+G_u)U^P+(-H_u^+)S^P$. Because our interest is in actuation at time $t$, we have $u(t)=U^f(1)$. This is computed at every time step. One should note the similarities with model predictive control (MPC), where the model is iteratively updated in conjunction with a cost minimizing control law at each time step, see \cite{Bordons2013}.

$H_u^+$ denotes the pseudo-inverse (\cite{Penrose1955}). A simple inverse amplifies high frequencies, yielding an impractical control law. Using a pseudo-inverse with non zero tolerance dampens high frequencies giving a smoother and hardware viable control law. In practice the tolerance level must be chosen such that actuation can follow the control law. Since actuator cannot achieve changes faster than 1 Hz, the tolerance level was chosen such that the impulse response control law did not exhibit fluctuations above 1 Hz, leading to a value of 2.5.

\begin{figure}
\centering
a) \includegraphics[width=0.46\textwidth]{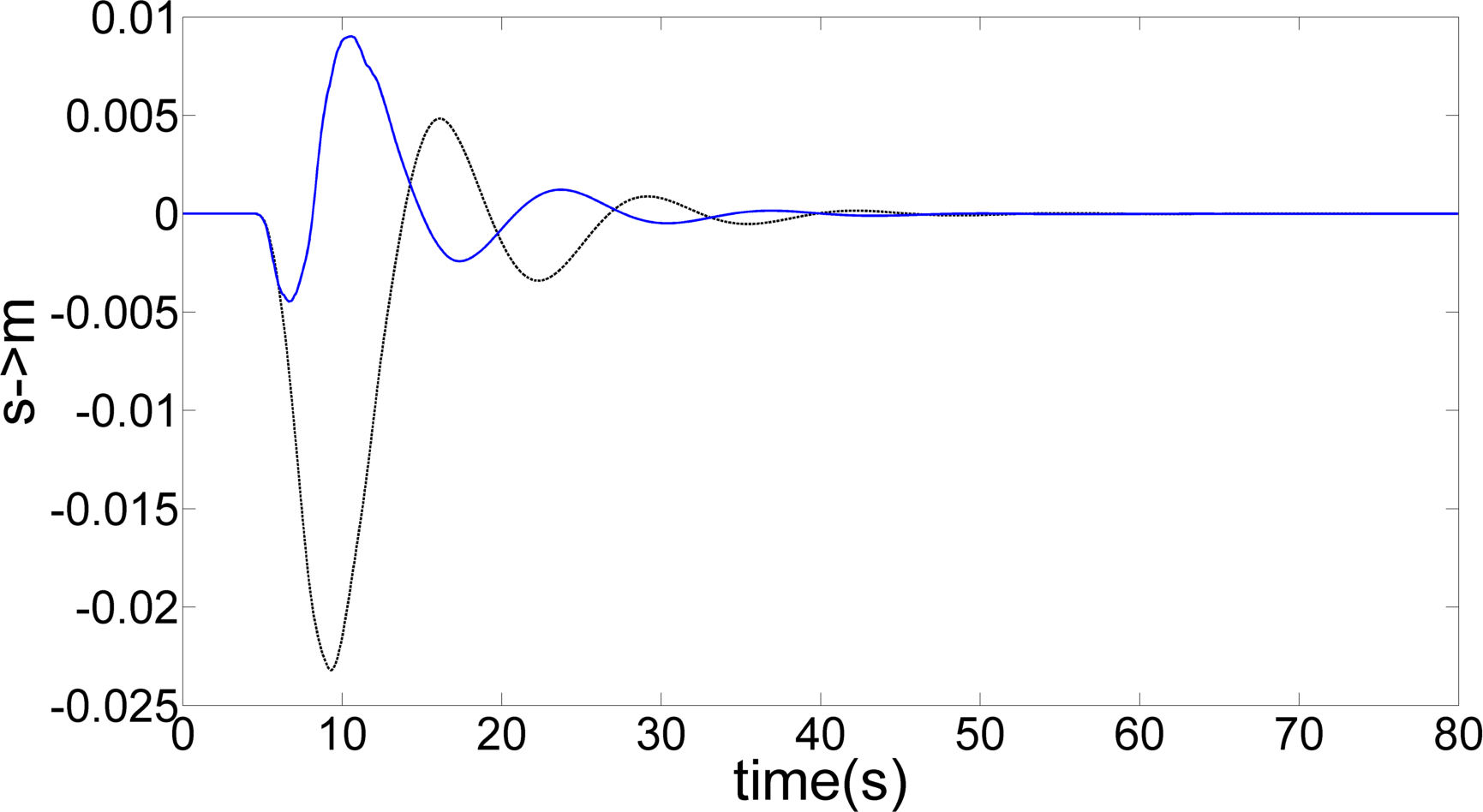}
b) \includegraphics[width=0.46\textwidth]{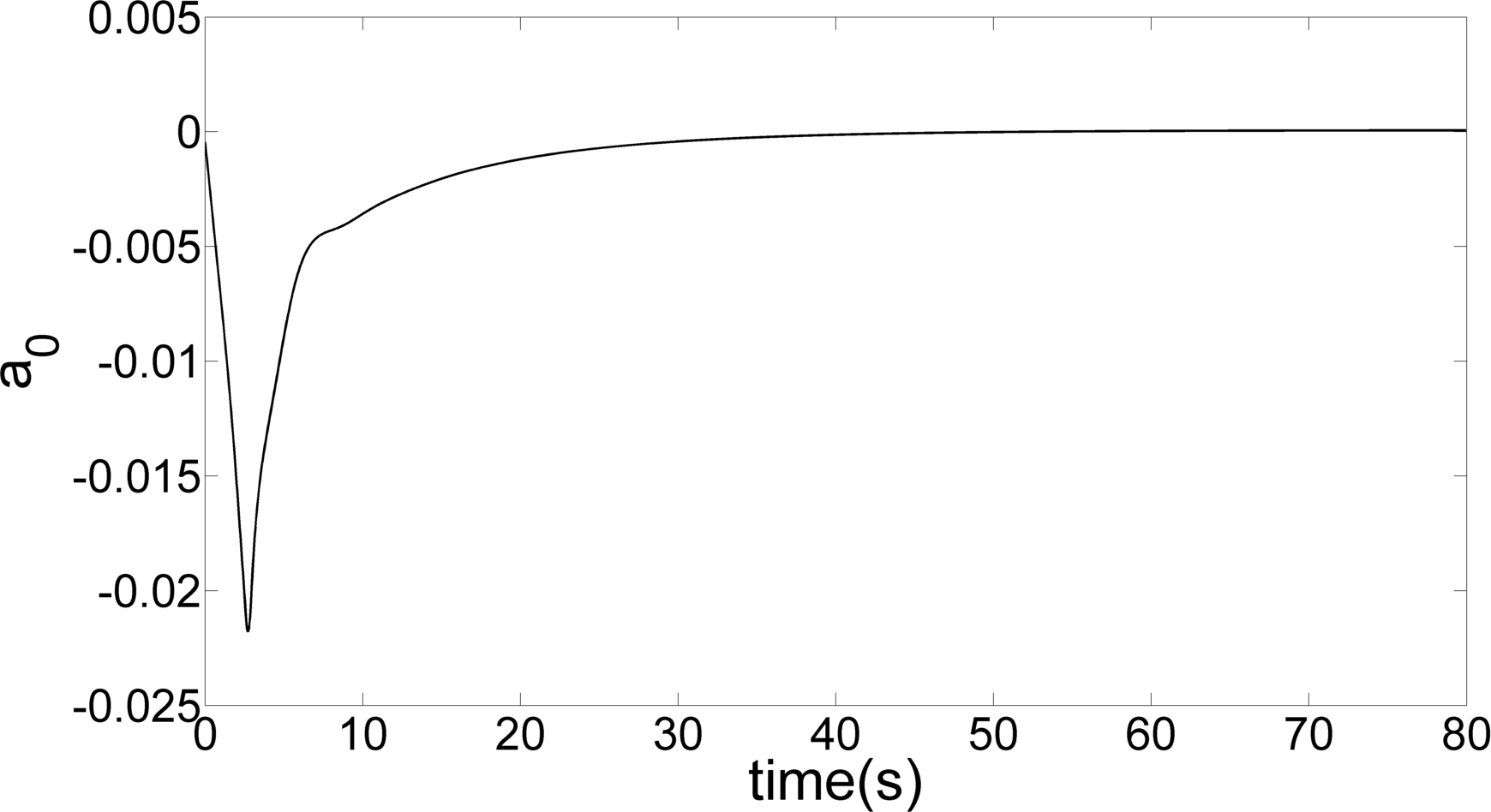}\\
\caption{a) Controlled (blue)  and uncontrolled (dotted black) response to an impulse in $s(t)$. b) Resulting control law $u(t)$.}
\label{fig:control_law}
\end{figure}

Figure \ref{fig:control_law}a compares the controlled and uncontrolled response of m(t) to an impulse in $s(t)$. Figure~\ref{fig:control_law}b shows the corresponding non dimensional control law $a_0 (t) = u_j (t) / U_{\infty}$. These figures show that while complete fluctuation negation is impossible, fluctuation damping is achievable. Such a control law will negate a portion of upstream disturbances. Furthermore since part of the perturbation will not have the chance to be further amplified in the shear layer this should result in noteworthy reduction in downstream TKE fluctuations. \cite{Sipp2012} found a far greater reduction for the impulse responses. One of the reasons for this is the location of the actuator, at the wall in our experiment, instead of in the bulk above the wall in the numerical simulation. The vortices created by the obstacle travel too far from the wall (approximately one step height) to be as successfully suppressed. 

\begin{figure}
\centering
\includegraphics[width=0.65\textwidth]{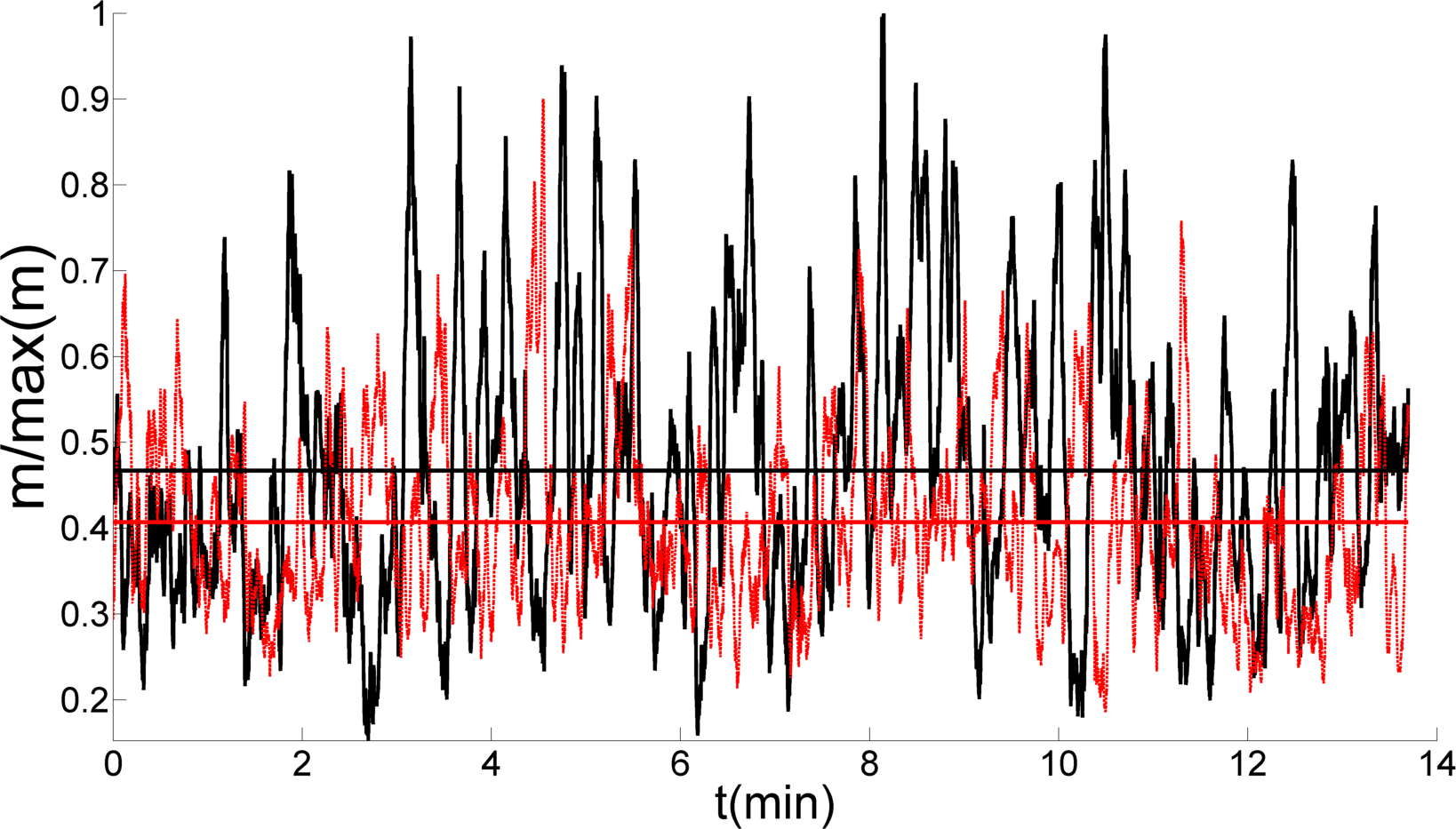}
\caption{Controlled (dotted red) and uncontrolled flow (thick black) outputs. Mean values are also displayed.}
\label{fig:control_results}
\end{figure}

\subsection{Control results}

\begin{figure}
\centering
a) \includegraphics[width=0.65\textwidth]{mean_TKE_2D_un}
b)\includegraphics[width=0.65\textwidth]{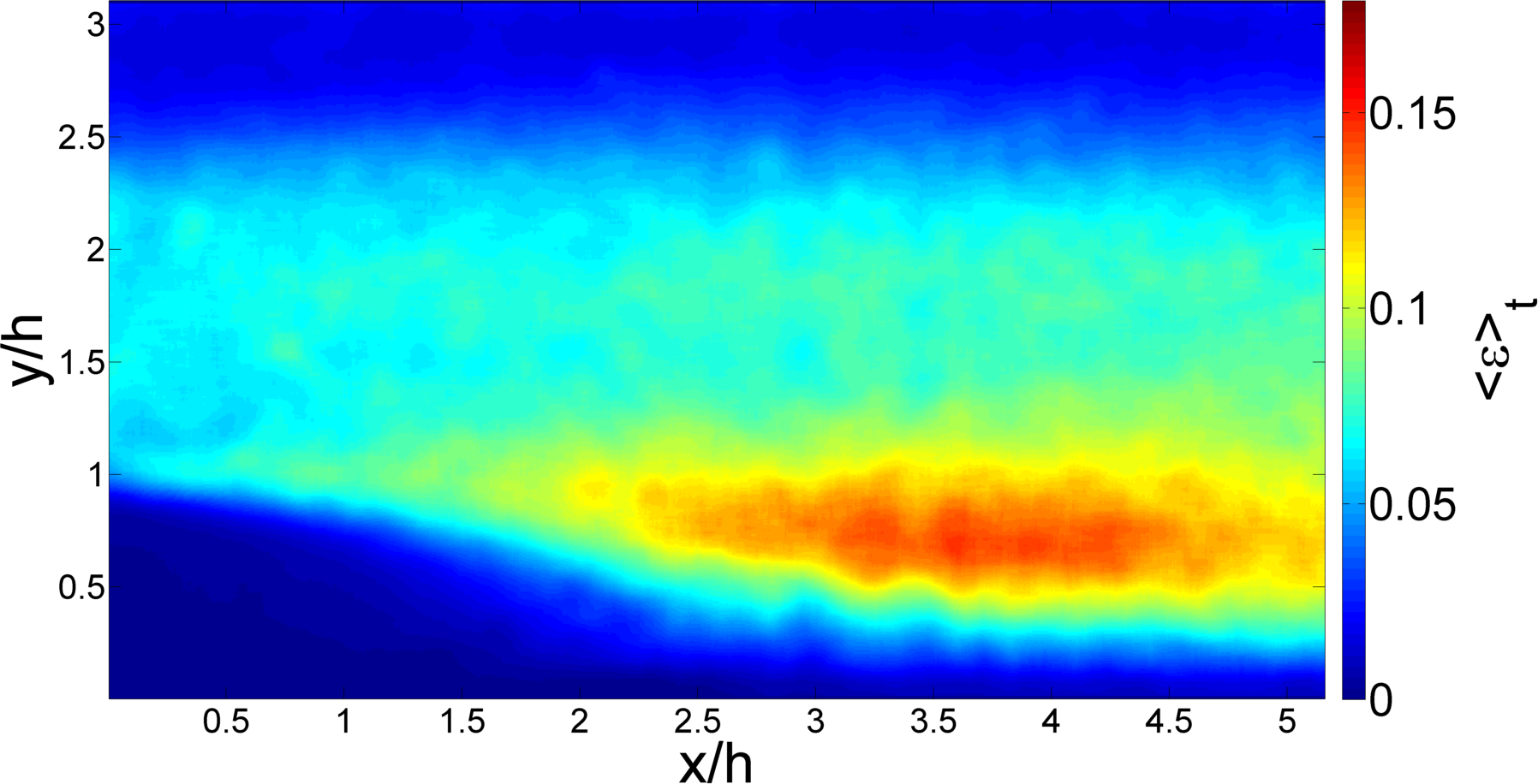}\\
\caption{Comparison of the time-averaged 2D TKE field  obtained for the uncontrolled (a) and controlled (b) flow.}
\label{fig:TKE_2D}
\end{figure}

\begin{figure}
\centering
a)\includegraphics[width=0.57\textwidth]{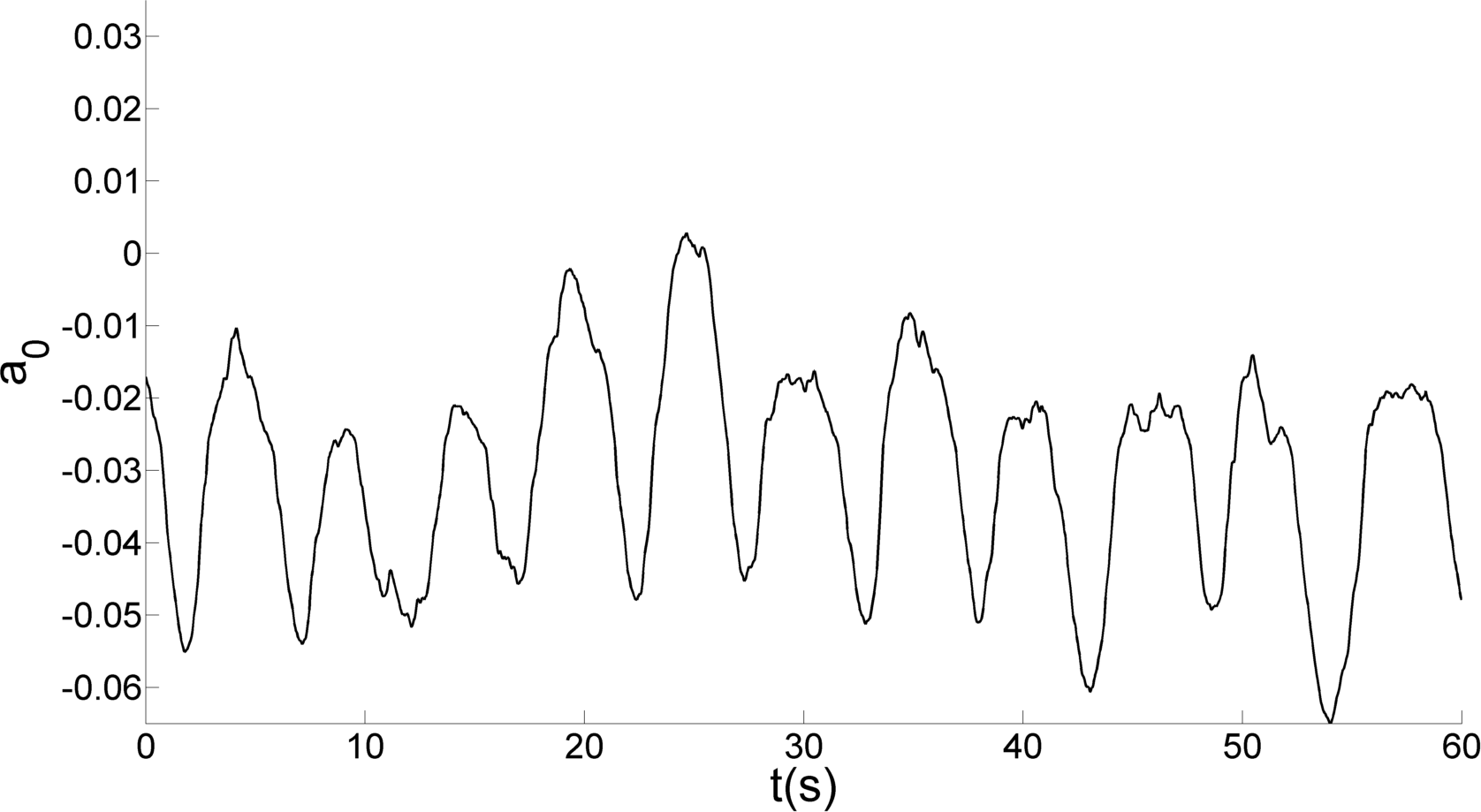}\\
b)\includegraphics[width=0.47\textwidth]{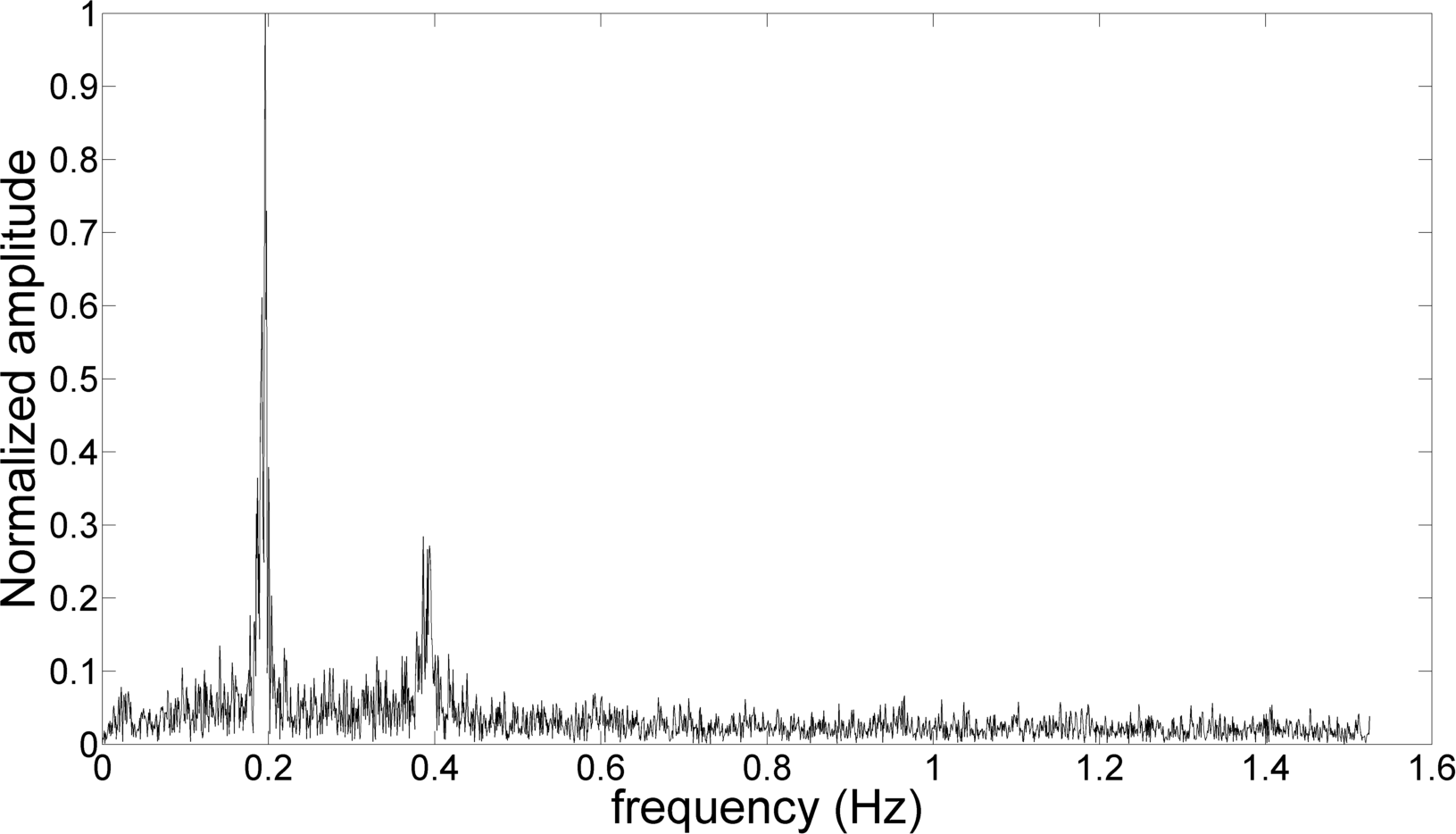}
c)\includegraphics[width=0.47\textwidth]{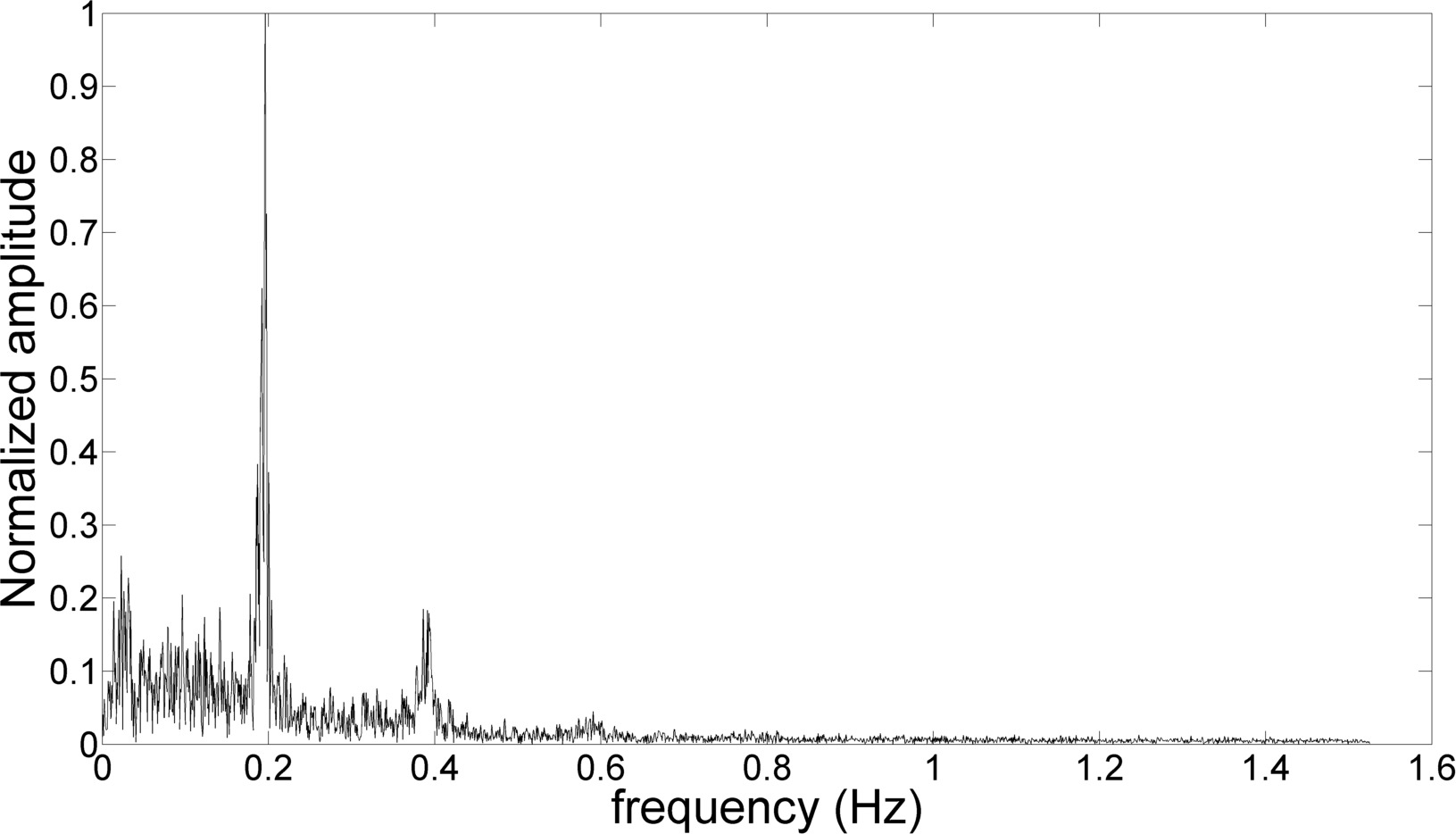}\\
\caption{a) Control law output $u(t)$ over one minute. b) Normalized frequency spectra for $s(t)$. c) Normalized frequency spectra for $u(t)$.}
\label{fig:freq_u}
\end{figure}

Figure~\ref{fig:control_results} shows a comparison between outputs for the controlled and uncontrolled flow. Comparison was done over 14 minutes (21000 iterations). The results clearly show a reduction in fluctuations for the controlled flow (-35 \%). Moreover a reduction in mean value is also observed (-15 \%). The mean value reduction is an added benefit of fluctuation reduction. Better performances could be expected when considering the impulse responses. Additional noise sources not accounted for by the upstream sensor are likely to be present in an experimental flow, contributing to degraded performance.
\\
Figure~\ref{fig:TKE_2D} shows the mean TKE field for the controlled and uncontrolled flows in the region of interest. The reduction in mean TKE is clear, as is a slight augmentation in recirculation size. Furthermore the effects of control are heterogeneous: while the TKE in the recirculation is mainly unaffected, the region of high TKE induced by the obstacle is successfully suppressed.
\\
Figure~\ref{fig:freq_u}a shows the non-dimensional control output sampled over one minute. One can see that the control signal is one of  periodic suction. Figures~\ref{fig:freq_u}b and \ref{fig:freq_u}c show the frequency spectra for $s(t)$ and $u(t)$. A double peak is present in both spectra for the same frequency. This explains the physical processes involved during control. An incoming vortex is detected as a spike in $s(t)$. The response is a sharp aspiration as shown by figure~\ref{fig:control_law}. Thus, the control is operating in opposition.

\section{Conclusion}
For the first time, an experimental implementation of a feed-forward control algorithm based on a ARMAX model was conducted on a backward-facing step flow. Results show the validity of such an approach. Nevertheless, to ensure successful implementation special care should be given to actuation, in particular to prevent contamination of the upstream sensor. Moreover, this approach is limited to the linear regime of the flow. \\
Analyzing impulse responses gives valuable insight into the flows controllability as well as the potential for success of the method. While these responses tell us full negation of upstream disturbances is impossible, the computed model was able to reliably predict flow responses and yield a control law able to reduce energy levels and fluctuations. Future work should involve span-wise sensors and actuators thus allowing span-wise heterogeneous disturbances to be controlled as proposed and evaluated numerically by \cite{Semeraro2011}.

\section{Acknowledgments}
The authors gratefully acknowledge the support of the DGA (Direction G\'en\'erale de l'Armement).
%\newpage
\bibliographystyle{jfm}
\bibliography{Bibliography}

\end{document}